\documentclass[onecolumn]{aa}

\usepackage[utf8]{inputenc}
\usepackage{amsfonts}
\usepackage{amssymb}
\usepackage{nccmath}
\usepackage{amsmath}
\usepackage{natbib}
\usepackage{graphicx}
\usepackage{txfonts}
\usepackage{textcomp}
\usepackage{subfigure}
\usepackage[pagebackref,colorlinks,citecolor=blue,urlcolor=blue,linkcolor=blue]{hyperref}
\usepackage{wasysym}
\usepackage{color}
\usepackage[mathscr]{euscript}
\usepackage{appendix}

\graphicspath{{figures/}}

\usepackage{siunitx}

\usepackage{xargs}                      
\usepackage[pdftex,dvipsnames]{xcolor}  
\usepackage[colorinlistoftodos,prependcaption,textsize=tiny]{todonotes}
\newcommandx{\unsure}[2][1=]{\todo[linecolor=red,backgroundcolor=red!25,bordercolor=red,#1]{#2}}
\newcommandx{\change}[2][1=]{\todo[linecolor=blue,backgroundcolor=blue!25,bordercolor=blue,#1]{#2}}
\newcommandx{\info}[2][1=]{\todo[linecolor=OliveGreen,backgroundcolor=OliveGreen!25,bordercolor=OliveGreen,#1]{#2}}
\newcommandx{\improvement}[2][1=]{\todo[linecolor=Plum,backgroundcolor=Plum!25,bordercolor=Plum,#1]{#2}}
\newcommandx{\thiswillnotshow}[2][1=]{\todo[disable,#1]{#2}}

\begin{document}

\title{Local monitoring of atmospheric transparency from the NASA MERRA-2 global assimilation system}

\author{
A.~Guyonnet\inst{\ref{inst1}}
\and 
S. Dagoret-Campagne\inst{\ref{inst2}}
\and 
N. Mondrik\inst{\ref{inst1}}
}

\institute{
Department of Physics, Harvard University, 17 Oxford Street, Cambridge, MA 02138, USA \label{inst1}
\and 
CNRS-IN2P3 LAL 91898 Orsay, France \label{inst2}
}

\titlerunning{MERRA-2 for ground based astronomy}

\offprints{aguyonnet\@@fas.harvard.edu}


\abstract{Ground-based astronomy has to correct astronomical observations from the impact of the atmospheric transparency and its variability.
The current objective of several observatories is to achieve a sub-percent level monitoring of atmospheric transmission. A promising approach has been to combine internal calibration of the observations with various external meteorological data sources, upon availability and depending on quality. In this paper we investigate the use of the NASA Modern-Era Retrospective Analysis for Research and Applications, version 2 (MERRA-2) which is a general circulation model (GCM) and data assimilation system that renders freely available for any given site, at any time, all the parameters constraining atmospheric transmission. This paper demonstrates the extraction of the relevant atmospheric parameters for optical astronomy at two sites: Mauna Kea in Hawaii and Cerro Tololo International Observatory in Chile. The temporal variability for the past eight years (annual, overnight and hourly), as well as the spatial gradients of ozone, precipitable water vapor, and aerosol optical depth is presented and their respective impacts on the atmospheric transparency is analyzed.}

\maketitle

\section*{Introduction}

The variability of atmospheric transparency above telescopes introduces a systematic uncertainty in their photometry. The traditional method to account for it is to normalize the observations against multi-epoch exposures of a set of stable stars (\citealt{2009A&A...506..999R}). The quality of the correction is a function of the type and number of reference stars within a given field of view, as well as the number of reference frames. The limits of this empirical method are twofold: the information is limited by the coarse resolution of the passbands, and by the evolution of the telescope throughput. The most successful methods (\citealt{2018AJ....155...41B}) mitigate these factors by adding external priors on several atmospheric parameters and demonstrate a stable broadband calibration at the level 5-6 permil. The systematic evaluation of the short term and long term  variability of these parameters is the main topic of this paper, along with their relative impact on the calibration of the atmospheric extinction.

Global assimilation systems developed by earth science experiments now routinely record the level of atmospheric constituents that translate into sub-percent variability of the atmospheric throughput. The avalanche of data has been organized in a comprehensive manner through reanalysis which combine observations using general circulation models to produce a best-guess reconstruction of the prior atmospheric state. Since these projects are central for climate researches, studies of atmospheric processes and evaluations of remote sensing systems, they have been under strong scrutiny and benefits from extensive analysis of consistency of the observations and the performance of assimilation methods.
The Modern-Era Retrospective Analysis for Research and Applications, version 2 (MERRA-2) (\citealt{randles} and references therein) is a NASA GMAO reanalysis project which assimilates satellites data from space agencies (NASA, ESA, Taiwan) as well as other external probes, such as AERONET, aircrafts and cruise ships onboard instrumentation (\citealt{doi:10.1175/JCLI-D-16-0758.1}). Its data product delivers all the atmospheric parameters needed to infer atmospheric extinction for any geographic coordinates since 1980 to present.

This paper demonstrates the extraction from  MERRA-2 data product of the optical transmission for two astronomical sites: Mauna Kea in Hawaii, and Cerro Tololo International Observatory (CTIO) in Chile. Section \ref{sec1} presents the results of the local vertical integration of longitudinally and latitudinally interpolated PWV, aerosol optical depth (AOD) and ozone column depth. Along with barometric pressure, the set of parameters allows to fully constrain radiative transfer simulations which deliver atmospheric transparency curves in the optical domain as a function of time and pointing: In section \ref{sec2}, the LibRadTran simulator (\citealt{mayer2005}, \citealt{1755-1315-28-1-012010}, \citealt{2015A&A...576A..77S}) is used to translate the variabilities of the atmospheric parameters into the variability of the atmospheric optical transmission. The uncertainties associated with the MERRA-2 data product are also propagated which offers a first assessment of both its quality and current limitations ounce applied to atmospheric transparency monitoring. The last section (§\ref{disc}) further discusses the perspective of using global earth observing systems in the context of ground based astronomy. In particular, the PWV and AOD levels recorded by above CTIO by two independent in situ observations ( \citealt{2014SPIE.9147E..6ZL}, \citealt{2018SPIE10704E..20C} ) are found to be within the MERRA-2 estimated uncertainties. In conclusion, the use of atmospheric transmission from MERRA-2 depends on the observational requirements: it can either be use as a self sufficient calibration, or, for the most demanding experiments, such as the Large Synoptic Survey Telescope (LSST), it could usefully supplement in-situ monitoring instruments.

\section{MERRA-2 atmospheric parameters extraction}
\label{sec1}

Modern-Era Retrospective analysis for Research and Applications Version 2 (MERRA-2) is a meteorological reanalysis undertaken by NASA’s Global Modeling and Assimilation Office (GMAO) which follows two primary objectives: 
(i) Place observations from NASA’s Earth Observing System (EOS) satellites into a climate context, (ii) Update MERRA system to include the most recent satellite data (\citealt{mccarthy}).
It is produced using the GEOS-5 atmospheric model and data assimilation system (\citealt{gmd-8-1339-2015}), a global mesoscale numerical simulations at 10 km resolution through 1.5 km resolution, and the three-dimensional variational data analysis (3DVAR) and Gridpoint Statistical Interpolation (GSI) meteorological analysis scheme (\citealt{wu}). The atmospheric model is resolved on a cubed-sphere grid with approximately 50 km horizontal resolution, and with 72 vertical layers from the surface to 0.01 hPa, on which an incremental analysis update procedure is performed every 6 hours.  It is then interpolated onto a 0.5 deg. latitude times 0.625 degree longitude horizontal grid for distribution to end users. Along with ozone and water, it also considers the aerosol fields in the reanalysis which is an important component of optical variability.

\subsection{Ozone}

The total ozone column observations are extracted from the Ozone Monitoring Instrument (OMI), onboard NASA’s EOS Aura satellite (\citealt{doi:10.1002/2014JD022493}), while the stratospheric profiles are extracted from the Microwave Limb Sounder instrument. The ozone field is computed from a chemical transport model powered by the GCM and the GEOS-5 atmospheric model. It is performed daily at four synoptic time, using 6 h model forecasts and observations within a ±3 h window of the analysis time. 
\cite{doi:10.1029/2007JD008802} discuss validation of the OMI daily total ozone against ground station measurements: On large space (continental) and time (months) scales, an offset of +0.4\% is found with OMI total ozone (OMI being higher), while aircraft measurements indicated -0.2\% offset with an RMS difference of 3\%. Assuming average OMI total ozone of 300 Dobson Unit (1 DU=0.01 mm of trace gas when the total column is compressed down to sea level at standard temperature and pressure), these numbers indicate a small offset of $\approx$1 DU and RMS difference of 9 DU, varying between 2 and 12, depending on the latitudes. For instance, 12 years of sonde-minus-analysis differences at Hohenpeissenberg show a mean residual of 1.43 DU and standard deviations of 8.1 DU (\citealt{doi:10.1002/2013JD020914}). 

\begin{figure}
\centering
\includegraphics[width=0.45\linewidth]{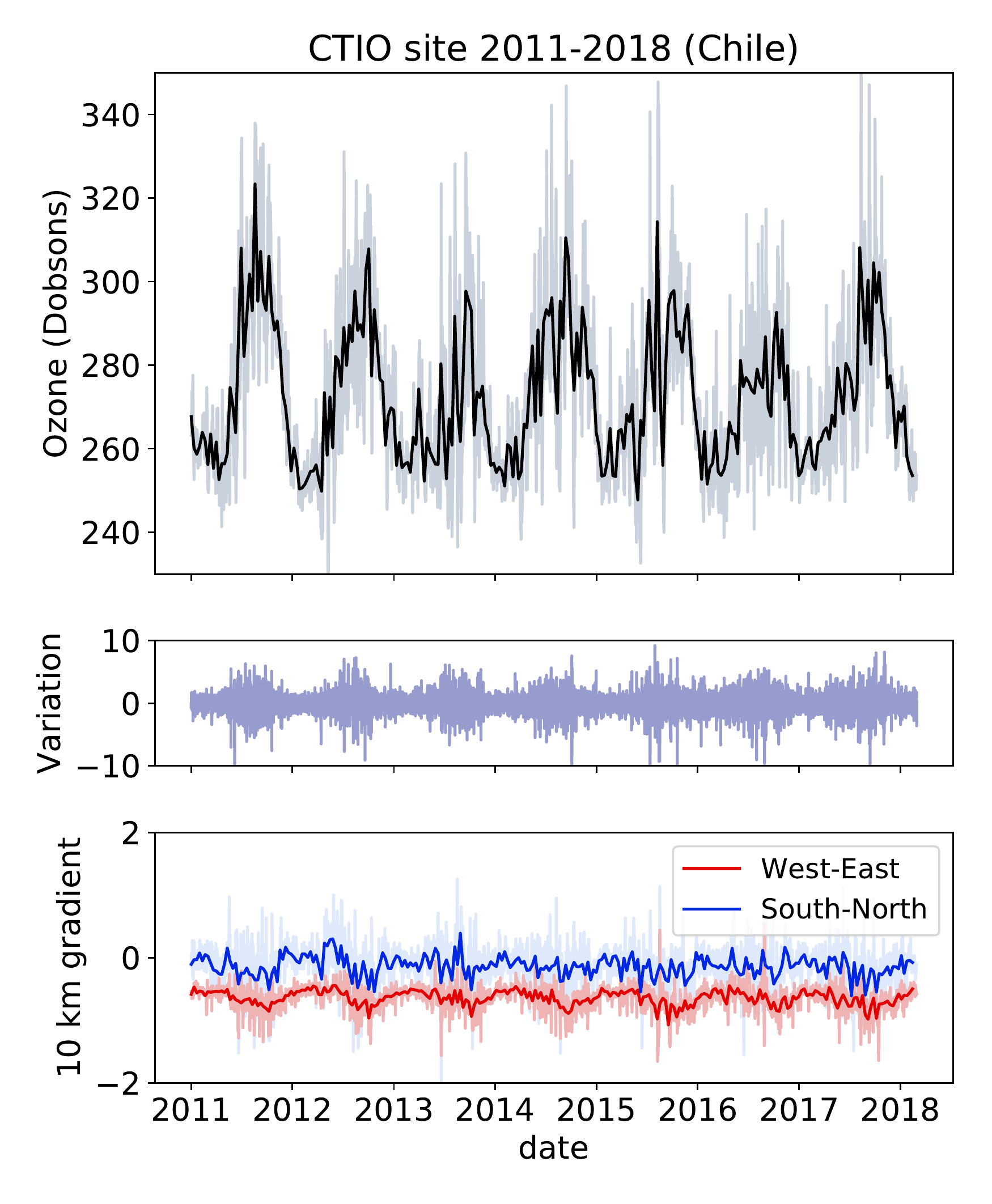}
\includegraphics[width=0.45\linewidth]{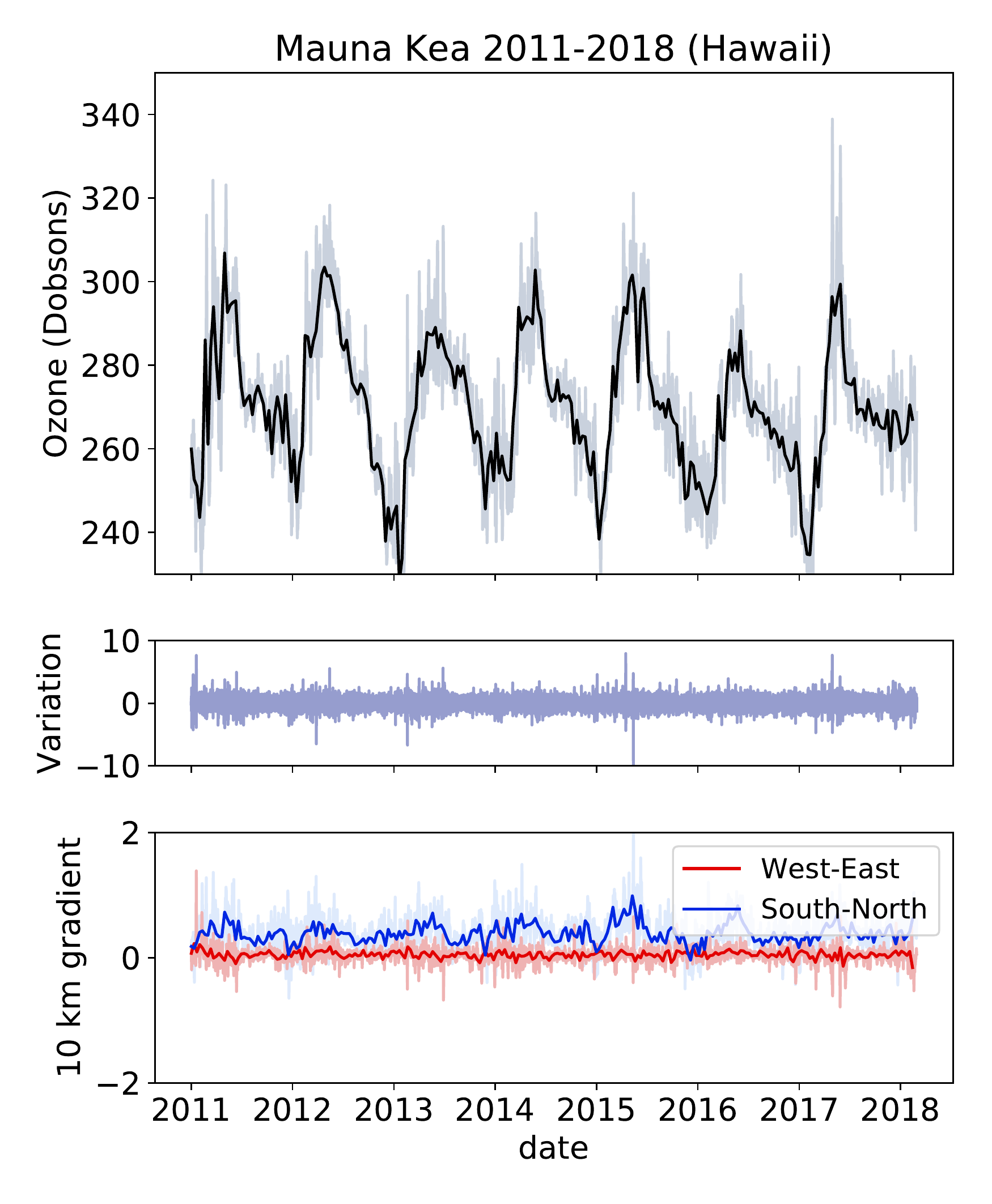}
\caption{Monitoring of the ozone depth above CTIO (top left panel) and Mauna Kea (top right panel) sites between 2011 and 2018. The ten days averaged time-series (black) above each sites exhibit a clear annual modulation, while the daily averaged values (faint grey) indicates significant daily variations on top of the annual modulation. The middle panels show the hourly derivative: Above CTIO, the variability is in the few Dobsons range and correlates with the mean, while the hourly variability above Mauna Kea is insignificant. The lower panels present the spatial gradients in the West-East direction (red) and South-North direction (blue). A steady $\sim$1 Dobsons West-East gradient is detected above CTIO.
 \label{fig:iqv}}
\end{figure}

Most of the ozone content lies in the stratosphere, between 20-30 km (profiles are shown in appendix \ref{fig:A_o3}). This is well above the tropopause,  
so the variability is driven rather by jet stream winds than by local meteorological events. The integrated value at a given time and location can be obtained from latitude, longitude and time interpolation of the MERRA-2 Single-Level Diagnostics table M2I1NXASM (\citealt{Bosilovich}).  Figure \ref{fig:iqv} presents the recorded values between 2011 and 2018 at the Mauna Kea (Hawaii) and CTIO (Chile) sites. The median value for the CTIO and Mauna Kea sites are respectively 268 and 270 DU with variability 46 and 45 DU (10 to 90 percentile). The modulation peaks in March-April at Hawaii and October-November in Chile. During spring seasons of both hemispheres, the hourly variability (central panels) can be as high as 5 DU. The bottom panels report on the West-East (in red) and South-North (in blue) gradients on a 10 km scale. The scale is a good compromise to analyze the axis symmetry of the constituent: for a $\pm$ 30° pointing around zenith this corresponds to an altitude of z$_{site}$ + 5.7 km. Above Mauna Kea, over the period, the mean South-North gradient is 0.4$\pm$0.3 DU/10km, and 0.05$\pm$0.14 DU/10km along West-East direction. The figures are -0.1$\pm$0.3 DU/10km (South-North) and -0.6$\pm$0.2 DU/10km (East-West) above CTIO. The combination of $z$-profile (figure \ref{fig:A_o3}) and gradients  (figure \ref{fig:iqv} bottom panels) allows to turn the $a priori$ axis-symmetric hypothesis that is often made in astronomy into numbers: Observations which lines of sights are 45° apart will result in 1 DU (S-N, Mauna Kea) and 3 DU (E-W, CTIO) variations. It translates into a 0.4\% and 1.2\% modulation of the atmospheric absorption at 600 nm.

\subsection{Precipitable Water Vapor}

\begin{figure}
\centering
\includegraphics[width=0.45\linewidth]{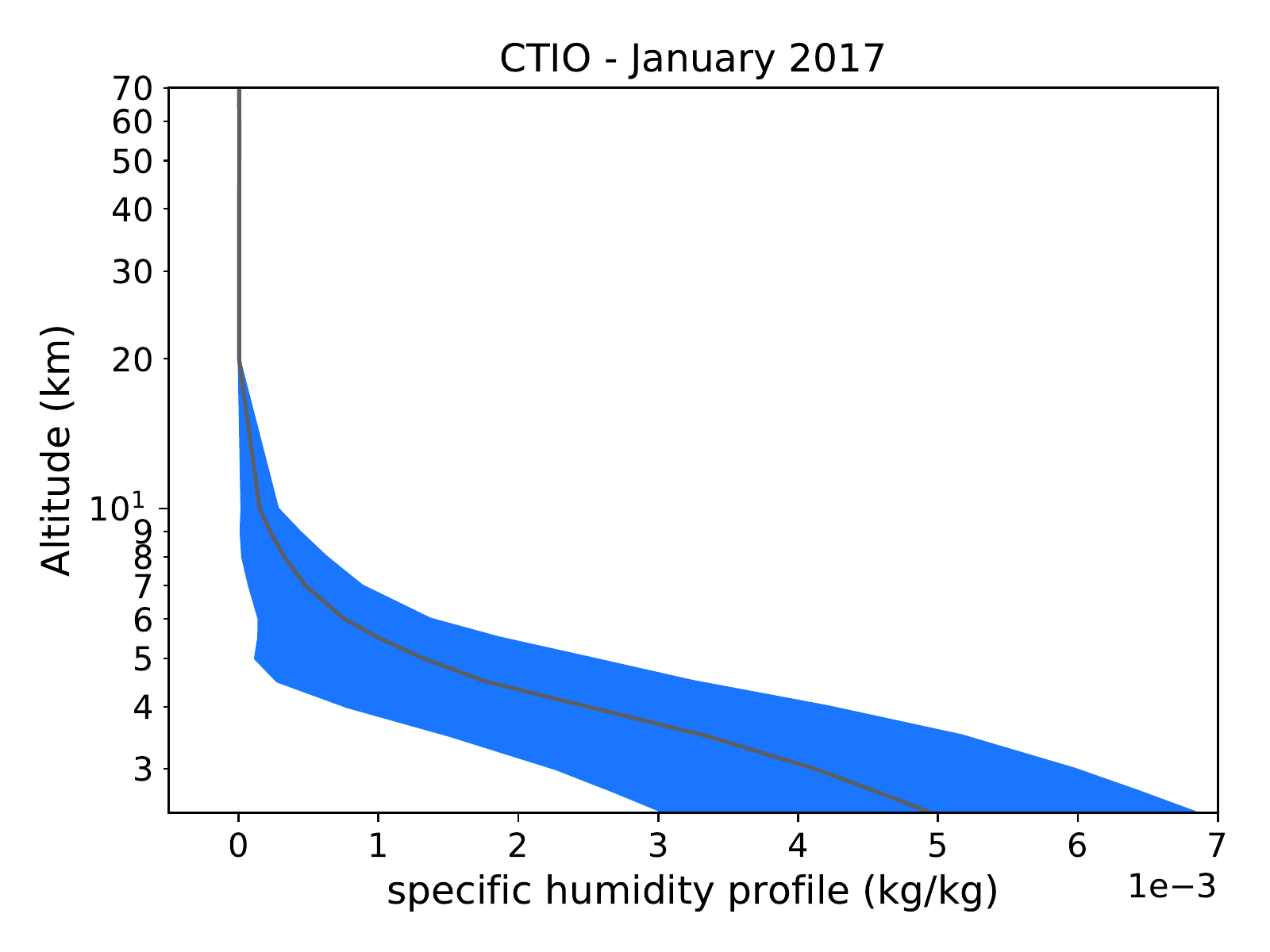}
\includegraphics[width=0.45\linewidth]{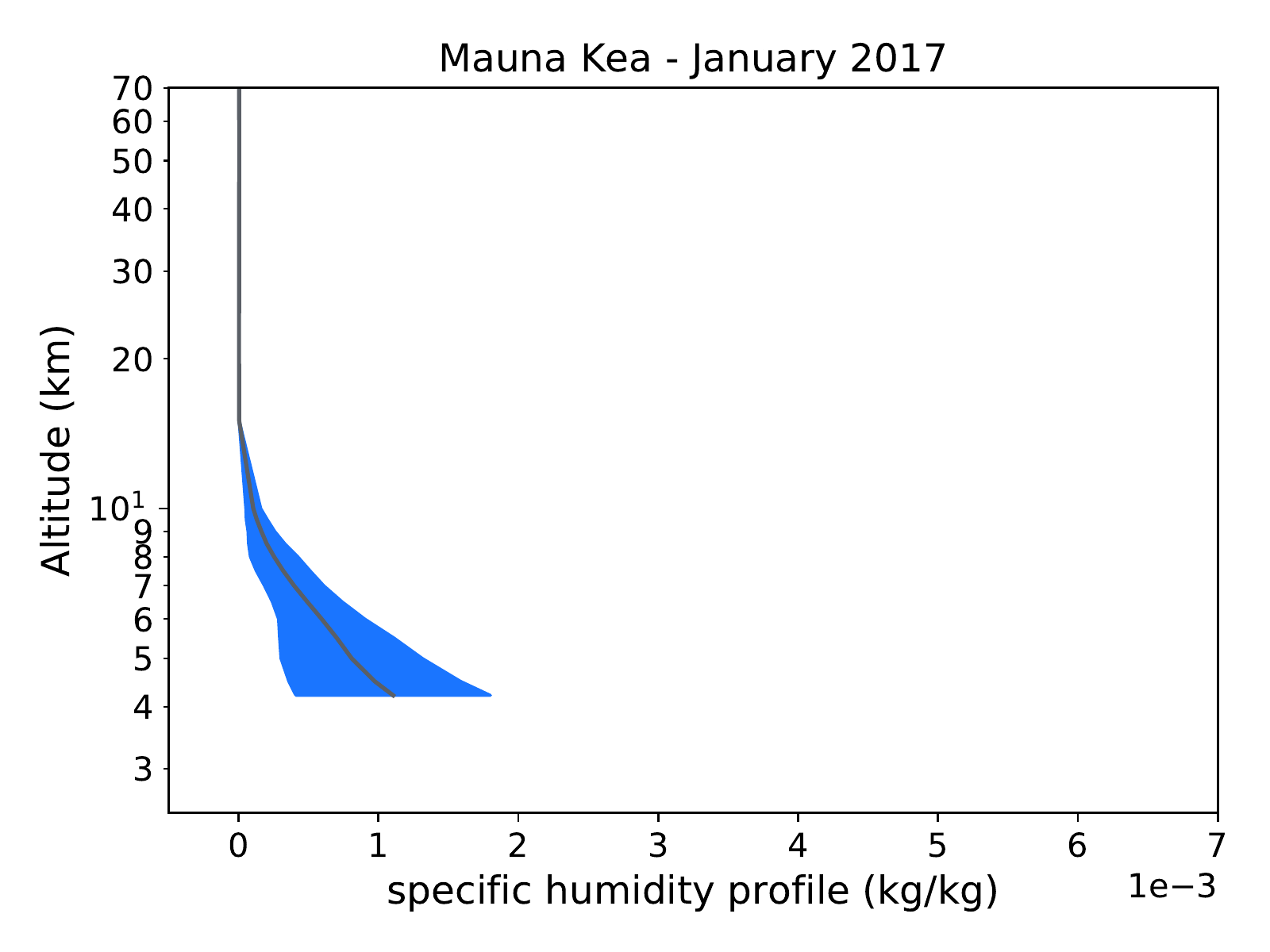}
\caption{Mean (grey lines) and 1 $\sigma$-RMS (blue area) interpolated specific humidity vertical profiles during January 2017 (one point per vertical bin, per 3 hours), from top of the atmosphere (70 km) to CTIO altitude (2300m, left) and Mauna Kea altitude (4200m, right). The altitude difference in between the sites makes the integrated humidity and its variability much larger above CTIO than above Mauna Kea. 
 \label{fig:3}}
\end{figure}

MERRA-2 monitoring of Precipitable Water Vapor (PWV) is obtained from assimilating a large amount of in situ and remote sensing observations into
an atmospheric general circulation model (\citealt{doi:10.1175/JCLI-D-16-0720.1}). An evaluation of the performance of the previous reanalysis (MERRA) in the context of abrupt orography is presented in \cite{wang} which report on a $\sim$10 \% bias from the comparison against ground-based GPS measurements at nine stations over the southern Tibetan Plateau from 2007 to 2013. PWV above a given telescope site can be obtained from the assimilated meteorological fields in the M2I3NVASM data product (\citealt{Bosilovich}) by vertically integrating the longitudinally and latitudinally interpolated specific humidity (weight of water vapor in the air per unit weight of air, in $kg/kg$). 
The PWV is the largest at the ground level. This is illustrated by figure \ref{fig:3} which shows the specific humidity vertical profiles above the CTIO and Mauna Kea telescope sites during the month of January 2017: The amount of water and its variability is larger at the altitude of the CTIO (2.3 km, left panel) than at the Mauna Kea summit (4.2 km, right panel). It become negligeable above 10 km.

\begin{figure}
\centering
\includegraphics[width=0.45\linewidth]{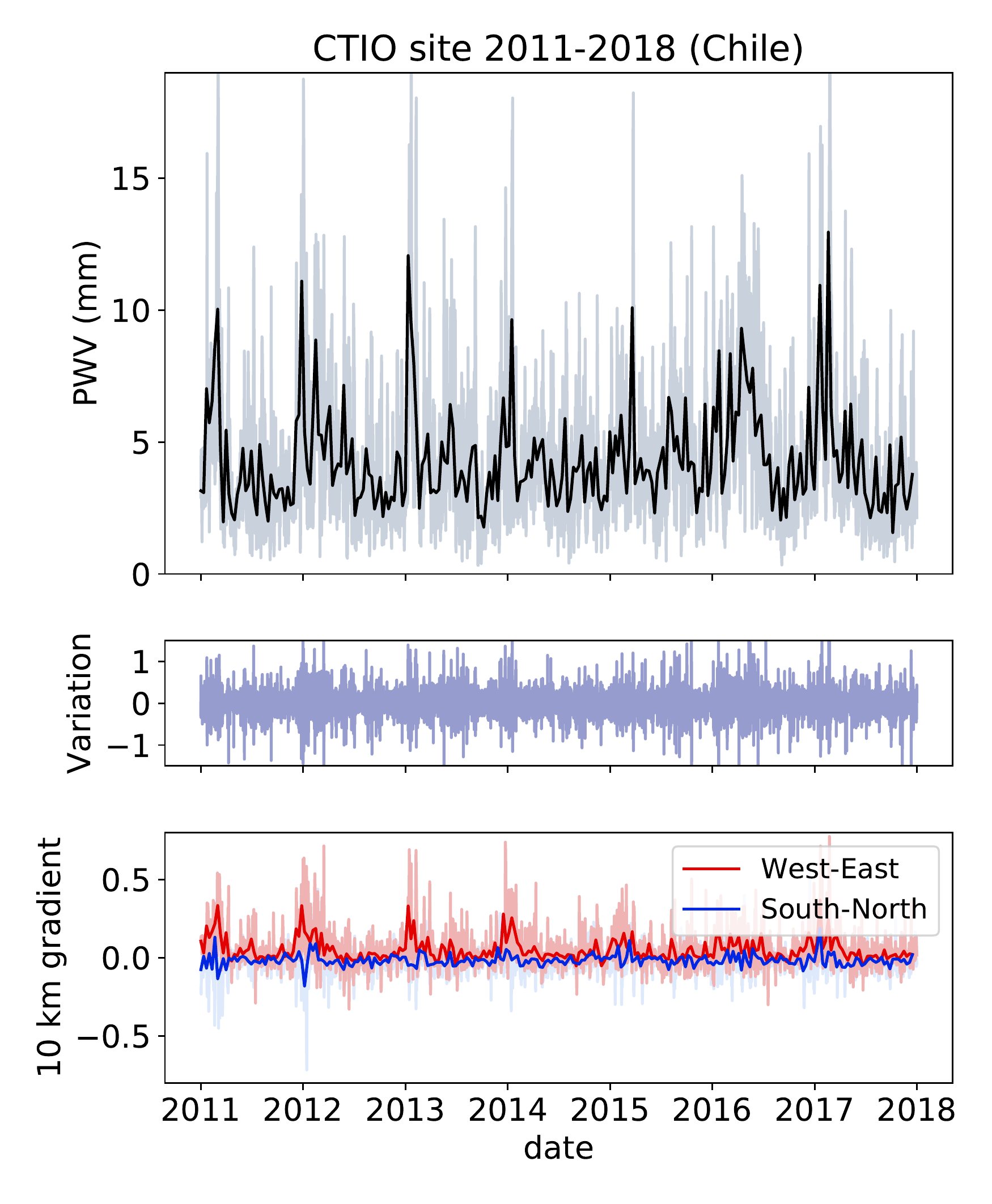}
\includegraphics[width=0.45\linewidth]{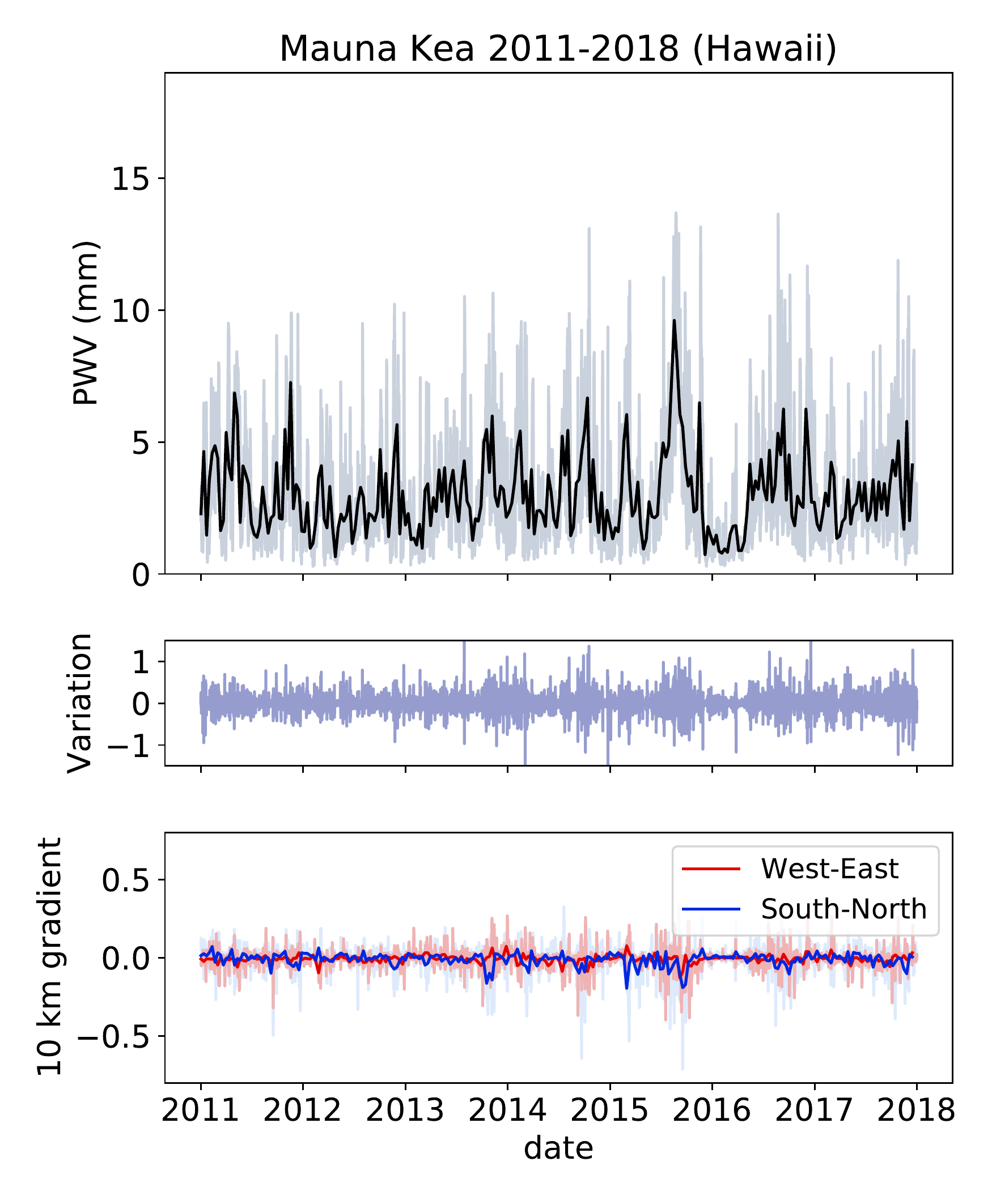}
\caption{Monitoring of the integrated specific humidity above CTIO (top left panel) and Mauna Kea (top right panel) sites between 2011 and 2018. The ten days averaged time-series (black) above each sites exhibit an annual modulation that is less pronounce than for the ozone. The variabilities of the daily averaged values (faint grey) are larger above CTIO than above Mauna Kea, which was expected from the vertical profiles shown figure \ref{fig:3}. Hourly derivatives are shown on the middle panels: this is sub-millimeter at both sites, and about twice smaller at Mauna Kea. The lower panels present the spatial gradients in the West-East direction (red) and South-North direction (blue). A West-East modulation is detected above CTIO.
 \label{fig:2}}
\end{figure}

The integrated specific humidity (in $kg/m^2$ or $mm$) from top of the atmosphere down to the telescope altitude level, as a function of time, is shown for both sites on the top two panels of figure \ref{fig:2}. MERRA-2 delivers continuous PWV records, so this includes cloudy nights. For the statistical analysis, only 10 to 90 percentile are kept. The CTIO site PWV median value for the period 2011-2018 is 3.60 mm with variability 6.54 mm. The Mauna Kea site median value is 2.26 mm with variability 5.33 mm. An annual modulation is also visible, but less pronounced than for the ozone: episodic large excursions events are spread throughout the year at both sites. The hourly variability (middle panels) is $\pm$0.15 mm above Mauna Kea and $\pm$0.23 mm above CTIO (figure \ref{fig:Apwv}, central panels). The erratic variation of PWV timeseries indicate that smooth overnight variation assumptions are not usually valid. The mean South-North gradient above the  Mauna Kea over the period is -0.01$\pm$0.08 mm/10km, and a West-East gradient at -0.006$\pm$0.06 mm/10km. The mean spatial asymmetries above CTIO are -0.01$\pm$0.10 mm/10km (S-N), and 0.05$\pm$0.12 mm/10km (W-E). A noticeable seasonal modulation of the West-East gradient above the CTIO is found (figure \ref{fig:2}, bottom left panel).

\subsection{Aerosols}

The MERRA-2 aerosol optical depth (AOD) is determined from observations by MODIS (Terra and Aqua satellites), NOAA Polar Operational Environmental Satellites (POES), NASA Earth Observing System (EOS) platforms, NASA ground-based observations, and AERONET \footnote{The AErosol RObotic NETwork \url{http:// aeronet.gsfc.nasa.gov/new_web/} is a federated global network of ground-based, automatic sun photometers that measure direct sun and sky radiances at several wavelengths reported uncertainty +/- 0.015 (\citealt{article}).}.
Aerosols are assimilated concurrently to the meteorological parameters using a radiatively coupled version of the  GOddard Chemistry, Aerosol, Radiation, and Transport model (GOCART) which treats the sources, sinks, and chemistry of 15 externally mixed aerosol mass mixing ratio tracers:

\begin{itemize}
\item Dust (5 non-interacting size bins with dry effective radii  0.64, 1.34, 2.32, 4.20 and 7.75 µm), 
\item Sea salt (5 non-interacting size bins with dry effective radii  0.08, 0.27, 1.05, 2.50 and 7.48 µm), 
\item Hydrophobic and hydrophilic black carbon (2 tracers, effective radius 0.04 µm),
\item Hydrophobic and hydrophilic organic carbon (2 tracers, effective radius 0.09 µm),
\item sulfate (SO$_4$, effective radius 0.16 µm).
\end{itemize}

 The model includes loss processes, including dry deposition, wet removal, convective scavenging and sedimentation (\citealt{buchard}). The dust and sea-salt particle size distribution is resolved across five non-interacting size bins each, with surface wind speed dependent emissions. The other components are prescribed from emissions inventories as well as chemical reactions. Sulfate (SO$_4$) and carbonaceous aerosol species have emissions principally from fossil fuel combustion, biomass burning, and bio-fuel consumption, with additional biogenic sources of particulate organic matter. The aerosols assimilation into GEOS-5 native 72 coordinate levels is performed at 8 synoptic times a day (0, 3, 6, 9, 12, 15, 18, and 21). The outputs are 15
three-dimensional fields of aerosol mass mixing ratio that can be integrated to deliver the Aerosol Optical Depth at 550 nm, following:

$$AOD_{550nm}^{site} = \sum_{AMR=1}^{15} \left( \sum_{z=top}^{z=site} \left(AMR(z) \times \beta_{550nm}(z)  \times  \Delta P(z)/g  \right) \right)$$

Where $AMR(z)$ is the Aerosol Mixing Ratio (kg kg$^{-1}$) of the 15 species, as a function of altitude $z$, tabulated in MERRA-2 tables M2I3NVAER, \cite{Bosilovich}. $\beta_{550nm}$ is the optical extinction coefficient derived from Mie theory and which depends both on the species and the relative humidity (see Supplementary Tables and Figures, \citealt{randles}). $\Delta P(z)$ and $g$ are respectively the barometric pressure in the bin $z$ and the gravitational acceleration. 
The relative humidity profile differs from the specific humidity profiles that are shown figure \ref{fig:3}: This is a function of the equilibrium vapor pressure, which evolves with temperature. As such, relative humidity is larger  in the tropopause, where temperature reach a local minimum, than on the ground (profiles above both telescope's sites are shown in appendix \ref{fig:10}). As a result, the optical depth of species with a large $\beta_{550nm}$ coefficient is driven by high altitude contents. The independent validation of the MERRA-2 aerosol products 
is rendered difficult because most of the global, readily available, space-borne and ground-based observations, are already included in the assimilation. However several tests are being reported in \cite{Randles2016} from which it is concluded that the bias between analyzed and observed AOD is generally within the $\pm$0.02 instrumental uncertainty.

\begin{figure}
\centering
\includegraphics[width=0.45\linewidth]{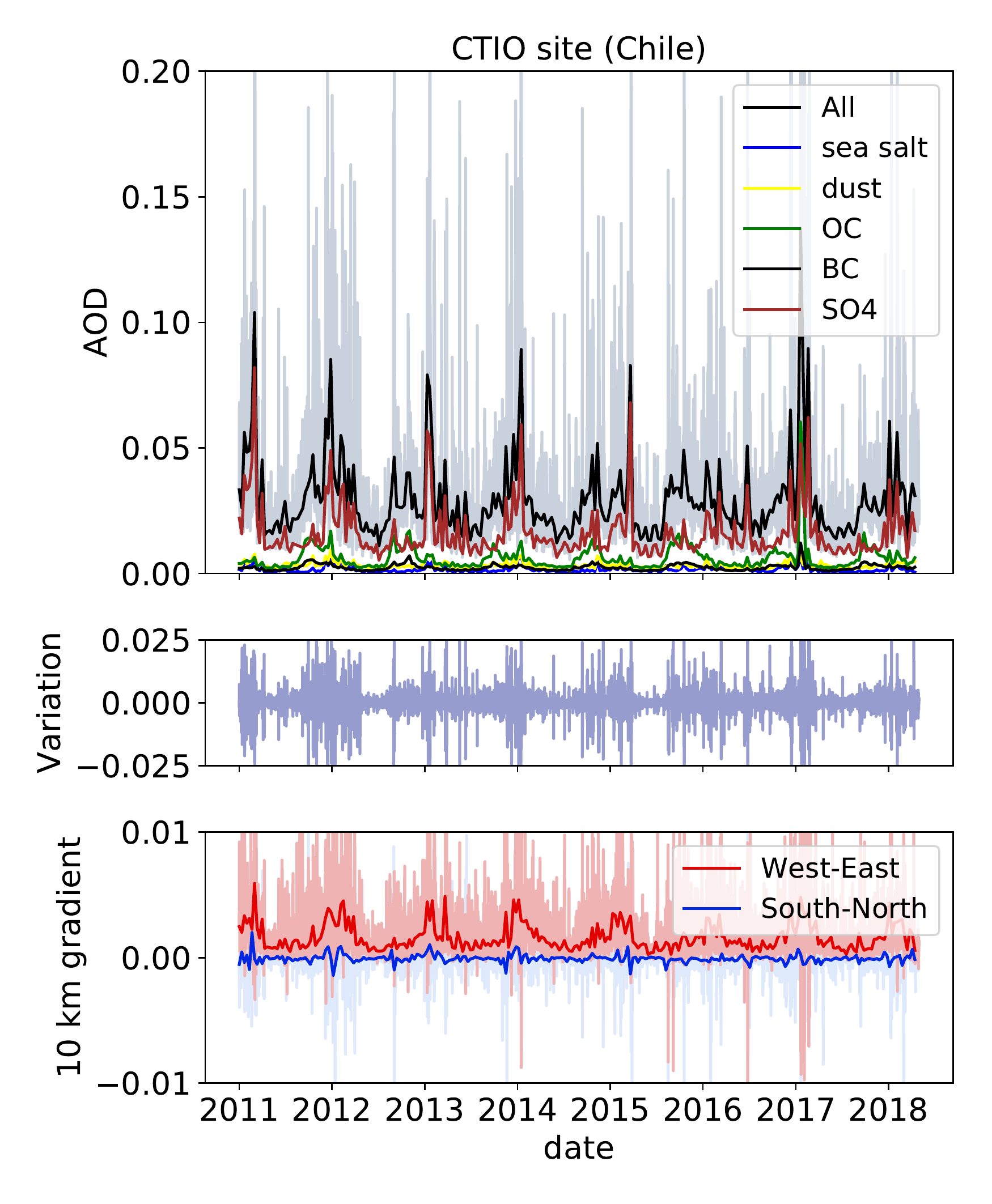}
\includegraphics[width=0.45\linewidth]{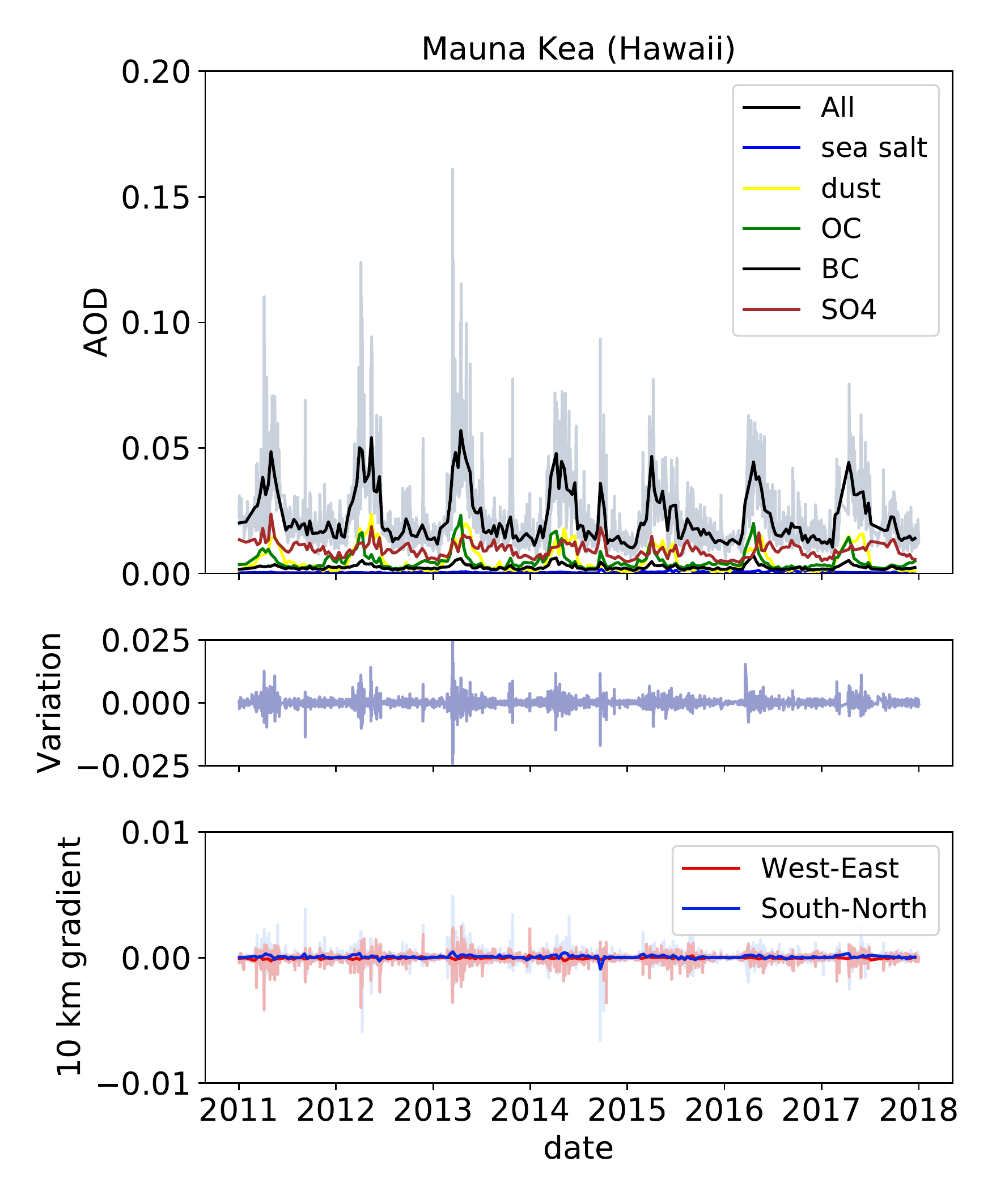}
\caption{Monitoring of the aerosol optical depth above CTIO (top left panel) and Mauna Kea (top right panel) sites between 2011 and 2018. The ten days averaged time-series (black) above each sites exhibit a clear annual modulation, while the daily averaged values (faint grey) indicates significant daily variations, especially above CTIO.
The panels also show the decomposition into five species: they all exhibit annual modulation, but spiking at different time of the year.
Middle panels show the hourly variation of AOD all together: the variabilities follow the annual modulations. The 10 km gradients (bottom panels) do not exhibit steady components, except for the West-East CTIO component that shows both an annual modulation and systematic bias toward the east.
 \label{fig:aod}}
\end{figure}

The MERRA-2 monitoring of the total AOD and its decomposition into species is presented figure \ref{fig:aod}. The AOD median value at CTIO for the period 2011-2018 is 0.023 with variability 0.038 (10 to 90 percentile) and about 20\% smaller at Mauna Kea, with 0.019$\pm$0.027 (dark lines, top panels). The contribution to the total AOD of the 5 dust bins is decreasingly strong: $\sim$0.0015 for the first,  $\sim$0.001 for the second, down to insignificant for the last. Second, third and forth sea salt bins are about one order of magnitude below, with few occasional burst at the level of a few permil (0.001-0.007). The hydrophilic organic carbon exhibits an annual modulation between 0.002 and 0.15 while hydrophobic organic carbon has insignificant contribution. Black carbon is about half the value of organic carbon and follows the same annual modulation. SO4 is the largest contributor to the optical attenuation with level between 0.01-0.02 and frequent bursts as high as 0.1. The sulfate mass extinction coefficient most strongly depends on the relative humidity ($\beta_{(RH=[0\%, 80\%, 95\%])}$= [3.15, 14.29, 22.53]). 
The amplitude of the hourly variation follows an annual pattern (middle panels), a feature that is more pronounced above Mauna Kea, thus indicating the predominance of higher altitude components. Regarding spatial variability, a noticable steady West-East gradient at 0.002$\pm$0.002/10km is visible above CTIO, also following an annual modulation. The gradients over Mauna Kea are negligible, at 7.0e-05$\pm$0.0003/10km (South-North) and -2e--05$\pm$0.0002/10km (West-East).

\section{Atmospheric transparency variability above CTIO and Mauna Kea sites}
\label{sec2}

The MERRA-2 atmospheric parameters can be translated into atmospheric transparency curves using a radiative transfer simulation. The ability of radiative transfer simulation to produce atmospheric transmission as a function of wavelength has been extensively tested (\citealt{mayer2005}, \citealt{1755-1315-28-1-012010}, \citealt{2015A&A...576A..77S}). In this work, the calculations of the atmospheric transmission is performed using the  Libradtran software package \footnote{http://www.libradtran.org/}. 

\begin{figure}
\centering
\includegraphics[width=0.45\linewidth]{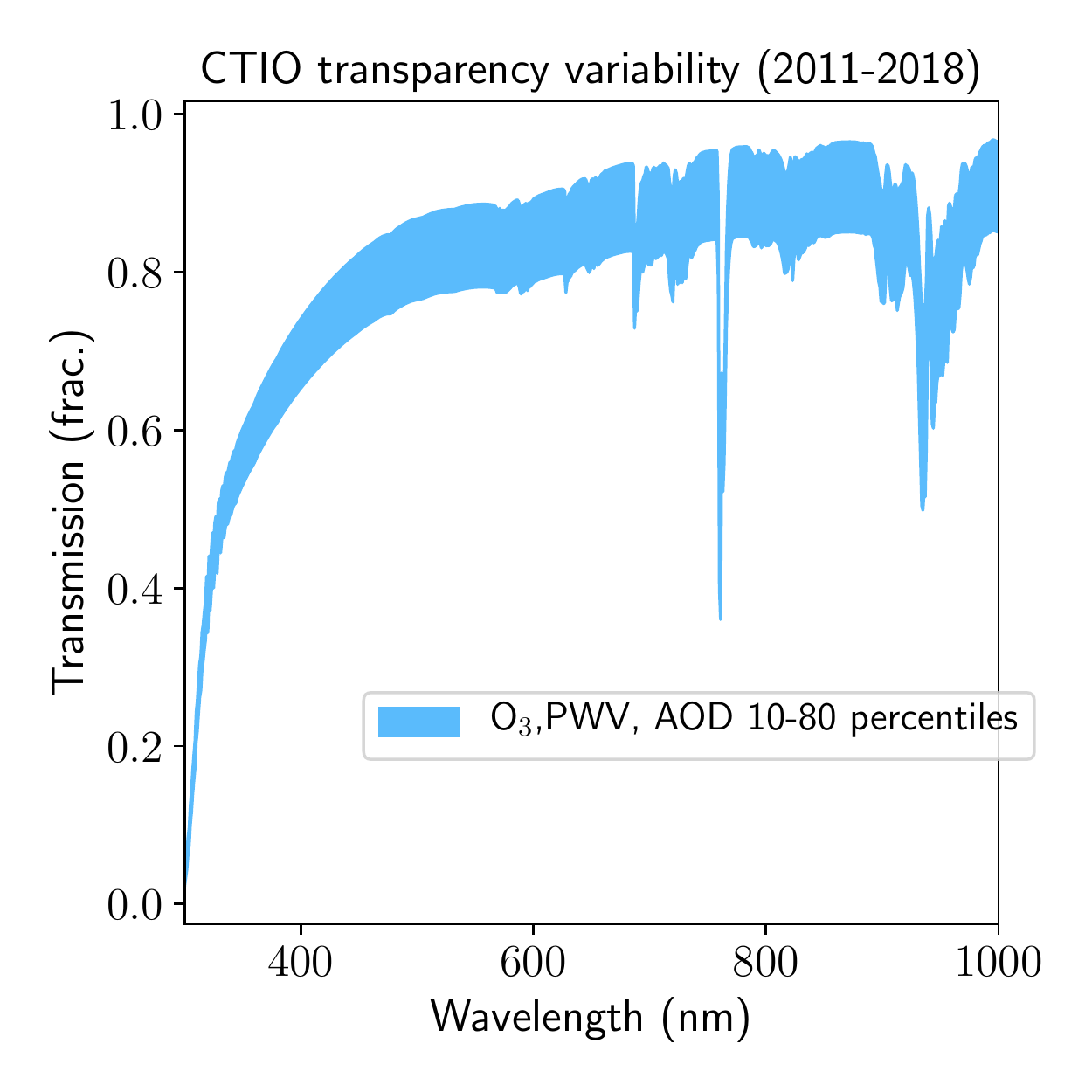}
\includegraphics[width=0.45\linewidth]{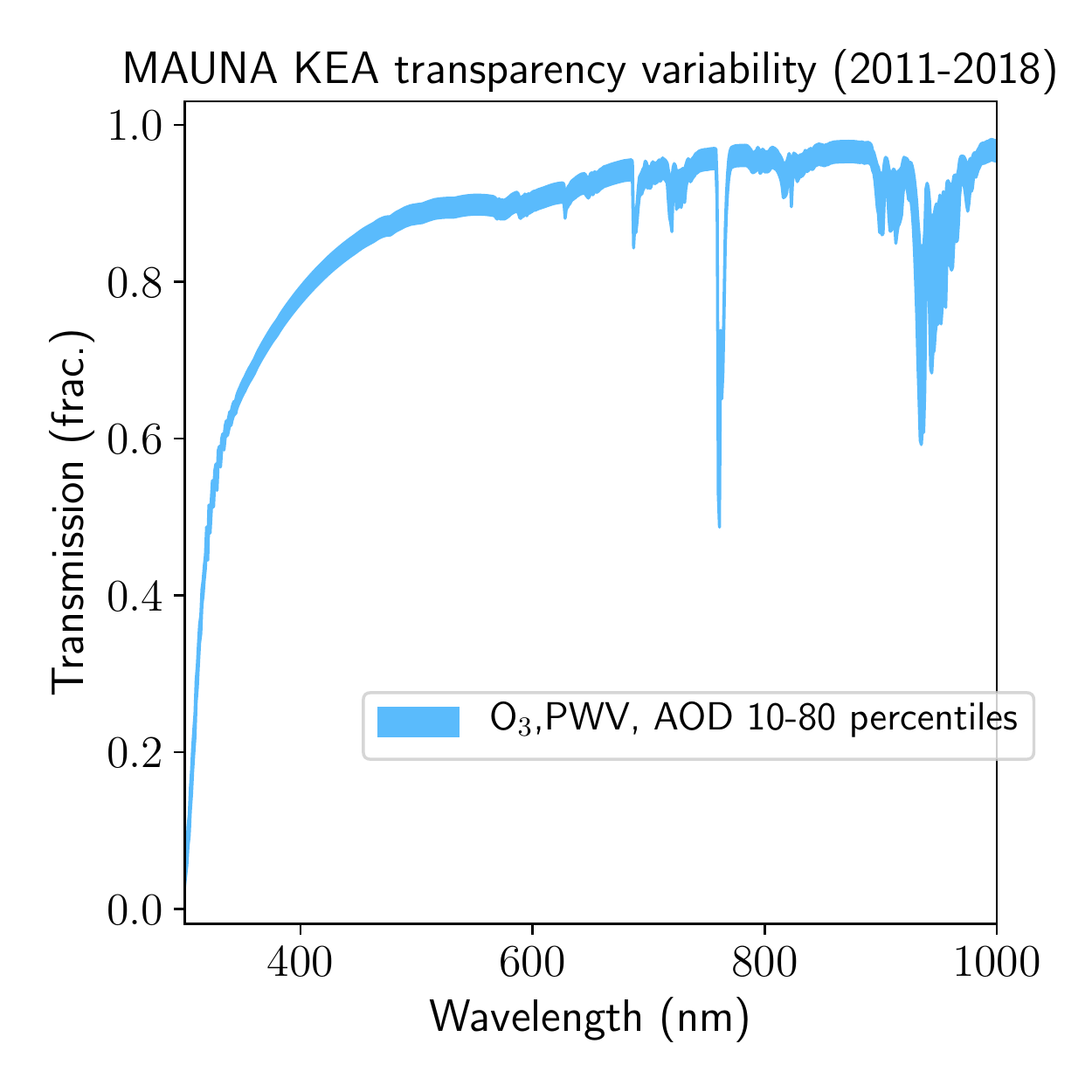}
\caption{Nominal transmission above CTIO site (left) and Mauna Kea (right) for the period 2011-2017. The variability of the transparency, corresponding to the 10-80 percentiles of the recorded O$_3$, PWV, and AOD parameters, is indicated by the blue area.
 \label{fig:abs}}
\end{figure}

\begin{figure}
\centering
\includegraphics[width=0.9\linewidth]{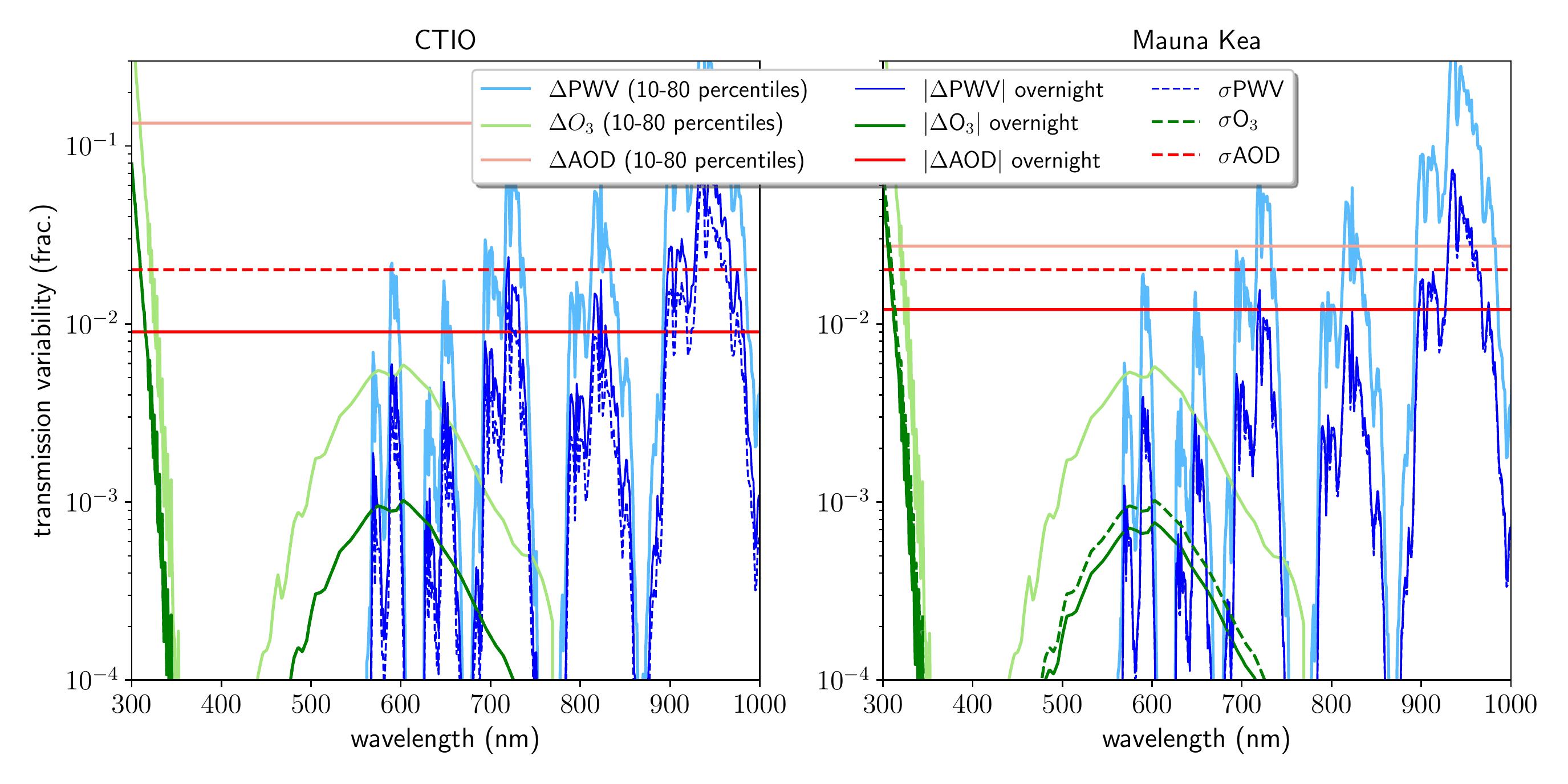}
\caption{Atmospheric attenuation variability above CTIO (left) and Mauna Kea (right) as a function of PWV (blue), O$_3$ (green) and AOD (red) parameters. Plain colors indicate the nominal modulation for the period 2011-2017 while  light colors inform on the typical overnight variabilities (|$\Delta_{max}^{overnight}$| 80 percentile). The propagation of the estimated MERRA-2 uncertainties are indicated by the dashed lines. The impact of ozone variability on the transmission is subpercent at both sites and well measured by MERRA-2. PWV variabilities are of the same order at both sites, with significant signatures in the red. The AOD estimate on the overnight variability is $\sim$1\%, which is a tenth of the annual modulation above CTIO, and a third of the Mauna Kea modulation.
 \label{fig:18}}
\end{figure}

The nominal transmission above both sites for the period 2011-2017 is shown figure \ref{fig:abs}. The variability of the transparency (blue area) corresponds to the 10 to 80 percentiles of the recorded O$_3$, PWV, and AOD parameters. It is observed that the overall transmission are of the same magnitude at both sites, but that the variability is about twice larger at CTIO (left panel). Part of the variability is induced by seasonal modulation (mainly from ozone and some aerosol species). An interesting information for astronomy is the analysis of the typical overnight variability: Figure \ref{fig:18} indicates the relative overnight modulation above both sites as a function of wavelength and depending on the parameter (darker blue for PWV, darker green for O$_3$, darker red for AOD).
At Mauna Kea, the typical overnight modulations (|$\Delta_{max}^{overnight}$| 80 percentile) are 1.13 mm for PWV, 5.9 DU for O$_3$ and 0.004  for AOD. The figures for CTIO are 1.84 mm for PWV, 7.9 DU for O$_3$ and 0.012 for AOD. These numbers are to be put in the perspective of 
 the nominal relative modulations over the entire period  (light colors), as well as the propagation of the estimated MERRA-2 uncertainties (dashed green corresponds to 8 DU for O$_3$, dashed blue corresponds to  1 mm for PWV and dashed red corresponds to 0.02 for AOD).

Each parameter has a specific signature on atmospheric transmission. The ozone column depth impacts atmospheric transparency curve in two distinct regions: the Huggins band, below 350 nm, and the Chappuis band, between 500 nm and 700 nm. The PWV imprints several features on the redder end of the optical spectrum, and the AOD modulates a continuous attenuation.
The ozone modulation has a subpercent impact on both the nominal and overnight atmospheric transmission variability, the PWV absorption features vary in the 1 to 10\% range on a nightly basis, while the AOD attenuation changes by $\sim$1\% overnight. In all cases, the MERRA-2 dataset is found to be reliable at reporting annual variations. Its level of precision reaches the range of the typical overnight modulation for the ozone and the PWV signals, while it is in between the nominal and overnight AOD attenuation variability, at $\sim$ 2\%.

\section{Discussion}
\label{disc}

High quality atmospheric parameter data sets with global coverage spanning multiple decades, produced by the earth science community, are of great value to other disciplines including astronomy. This work is a first step in assessing how this could benefit ground based optical astronomy in its need to correct observations of astrophysical scenes from the atmospheric transmission. The information that is of potential interest to astronomers is extracted in three steps: the MERRA-2 3-D tables are interpolated in latitude and longitude and integrated in altitude at the coordinates of the Mauna Kea summit in Hawaii, and the CTIO site in Chilean Andes. The hourly time-series between 2011 and 2018 of the atmospheric parameters are translated into atmospheric extinctions using the LibRadTran radiative transfer code, along with a propagation of the uncertainties estimated from the reference publications.

The result of this work is an analysis of the variability of the atmospheric optical transmission function. A first information obtained from the long term time-series is a picture of both seasonal and circadian modulations of transparency above the observatories. The uncertainties associated with the measurements are in the 2-4\% range, in between the amplitudes of the seasonal and circadian modulations and well above the hourly variations. The error budget can be further analyzed by looking at the relative contribution of its components with respect to their spatial distributions: the combination of $z$-profiles and latitudinal and longitudinal gradients shows that the variability is smoother for high altitude parameters (such as ozone and some aerosols) than components that are mostly distributed nearer to the ground layer (such as precipitable water and other aerosols) and which are also influenced by meteorological events on local scales. Given that there is a large contribution of satellite observations into the MERRA-2 data product, which are most notably effective at monitoring high altitude molecules, astronomers could likely benefit from using the long term time series of these components. In summary, the resolution of the MERRA-2 grid is a limitation mostly for the low altitude parameters (PWV and some aerosols) and not so much for high altitude constituents (ozone and some other aerosol species).

The relevance of MERRA-2 for ground based astronomy is both a function of the astronomical site location and the scientific use of the observations. The correlation seen between spatial distribution and annual signal modulation suggests that the higher the site, the most relevant the MERRA-2 monitoring. The analysis also brings a quantitative assessment of the axis-symmetry and smooth variation hypothesis that is often assumed when reducing astronomical observations: it is found that the validity of the hypothesis could be both a function of the altitude of the telescope site and of the atmospheric components. For ozone, that is mostly present in higher layers but with rather small horizontal gradients, the axis-symmetry and smooth temporal variation assumption seems valid down to a few per-mil.  For water, which accumulation and variability is the highest nearer to the ground, the validity of the hypothesis strongly depends on the altitude of the observing site. Lastly, aerosols are complex because the quality of MERRA-2 measurement is species dependent, which in turn is both location and altitude dependent. Site specific studies would therefore be needed. As a matter of fact, usefulness should be apprehended from two perspectives: (i) the site location and (ii) the calibration requirements.

\begin{figure}
\centering
\includegraphics[width=0.45\linewidth]{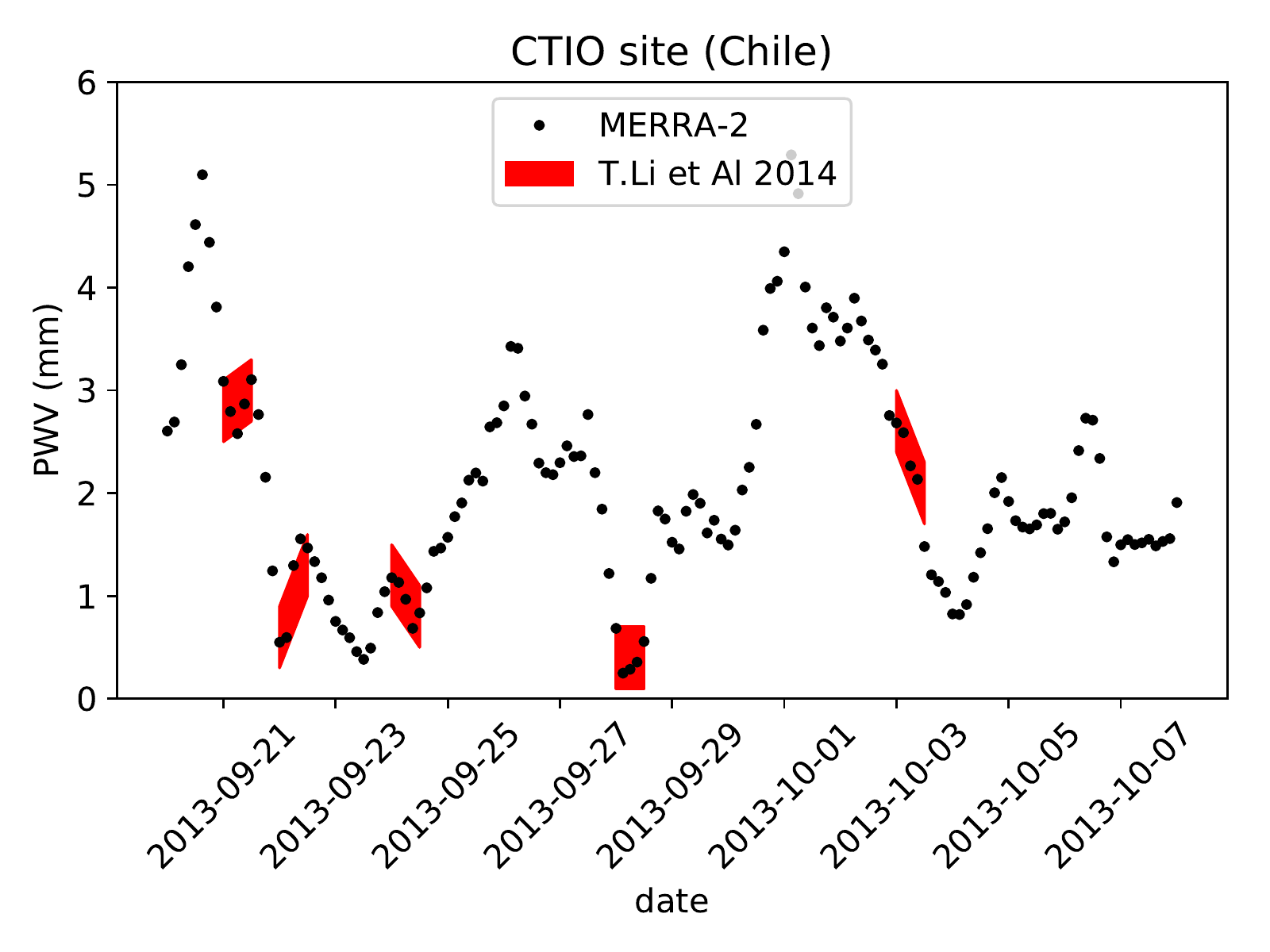}
\includegraphics[width=0.45\linewidth]{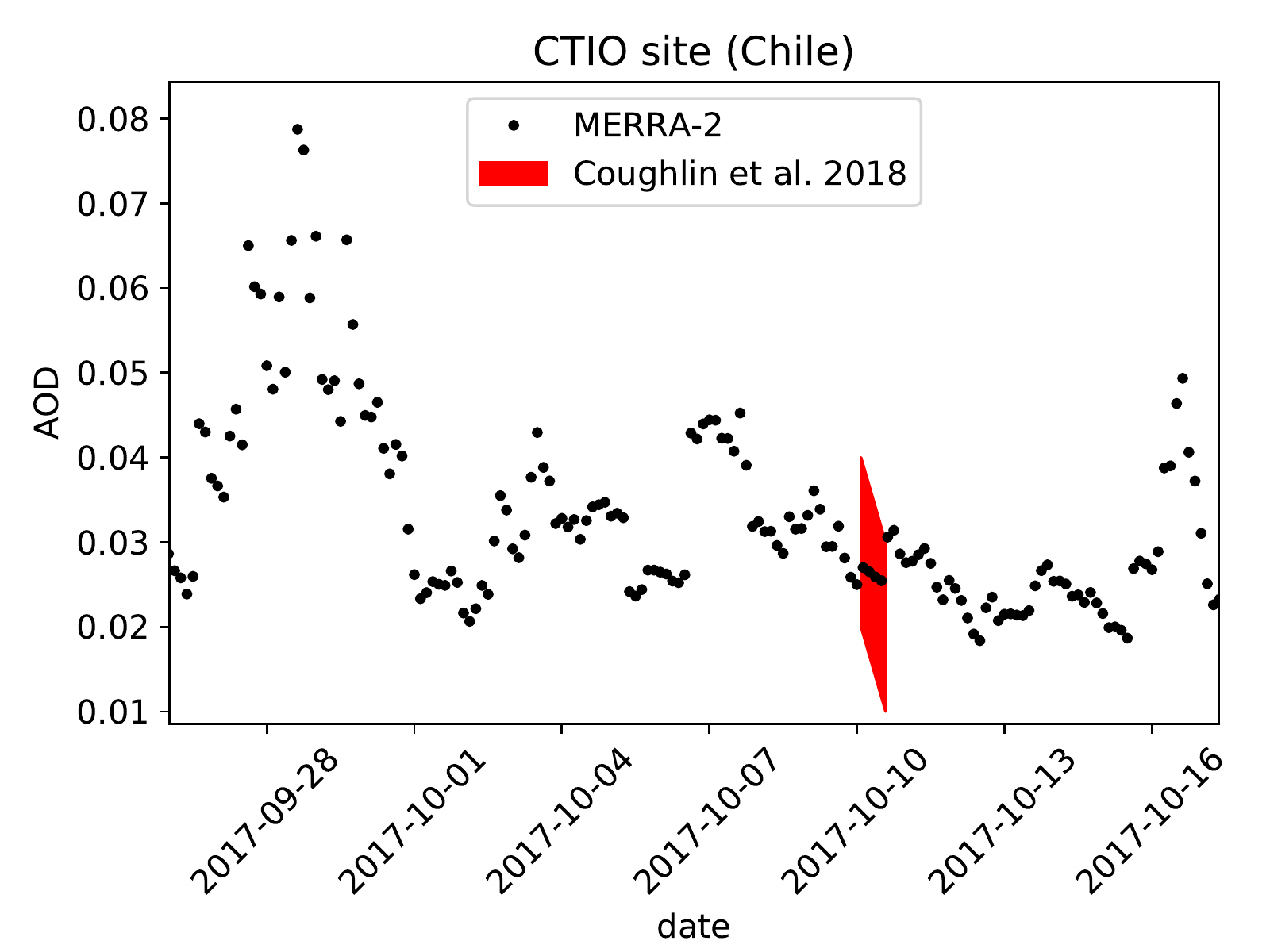}
\caption{Comparison of MERRA-2 monitoring (black points) with in-situ observations from Li et al 2014 (left) and from Coughlin et al 2018 (right) publications. The Li et Al. 2014 campaign measured PWV above CTIO using GPS and broadband photometry in Autumn 2013 (red polygons) while Coughlin et al 2018 measured AOD form spectrography of stars as a function of airmass in Autumn 2017. Both observations agree with MERRA-2 monitoring within the estimated uncertainties.
 \label{fig:19}}
\end{figure}

MERRA-2 can be compared with a myriad of local observations. For instance, \cite{2014SPIE.9147E..6ZL} measured the PWV level above CTIO in Autumn 2012 and 2013 from repeated narrow-band photometry of standard stars. The reading of the 2013 run (figure 5 of the publication) is plotted on figure \ref{fig:19} (left panel, red polygons) against the MERRA-2 time series at that time (black points). The two compare well and are consistent with both the 1 mm MERRA-2 uncertainty assumption from its literature and the 0.6 mm uncertainty reported by \cite{2014SPIE.9147E..6ZL}. A second comparison, on figure \ref{fig:19} right panel, shows the direct observation by \cite{2018SPIE10704E..20C} of the AOD above CTIO in October 10, 2017. AOD was measured from fitting the variability of the spectrograms of a reference star as a function of airmass during a photometric night. The $\sim$0.025 AOD (red polygon) that is reported is also in agreement with the level estimated by MERRA-2 (black points). These comparisons are encouraging and might indicate that the MERRA-2 precision could be slightly better than expected at this site. In the future, several aspects could be further investigated: The MERRA-2 time series and error modes could be compared against another assimilation system, such as ERA-5\footnote{ERA5 provides hourly estimates of atmospheric variables on a 30km grid and 137 levels from the surface up to a height of 80km. It also includes information about uncertainties at reduced spatial and temporal resolutions.}. The reliability of local interpolation of 3-D variables from global data in the context of abrupt orography could also be assessed by combining local probes with a mesoscale weather forecast algorithm to control the assimilated profiles.  Such forecast algorithms have already bene used successfully by \cite{2018MNRAS.477.5477P} at La Palma Roque de los Muchachos Observatory to predict PWV content at millimeter level precision and sub-millimeter accuracy. Lastly, comparisons with direct observations on more extended time span are needed, and could be easily achieved at these two sites given the many past surveys (such as the Dark Energy Survey or the SuperNovae Factory) and future surveys, such as the LSST.

The LSST (\citealt{Ivezi__2019}) is a next generation optical survey at CTIO site that will be targeting a per-mil level photometric calibration, a requirement driven by the objective to improve the precision of the current measurement of dark energy by a factor of 5, to reach a percent level (\citealt{2012arXiv1211.0310L}). The MERRA-2 reanalayis fields above the CTIO site can be used to estimate the expected residuals of the observations due to the atmospheric variability. The table \ref{tab:1} presents the residuals (in millimagnitude) for the u, g, r, i, z, y bands \footnote{from \url{https://www.lsst.org/scientists/keynumbers}}. They are computed following:
$$ m_i^b - m_j^b = - 2.5 log (\phi_i^b / \phi_j^b)  $$
Where the magnitude difference $m_i^b - m_j^b$ in a band $b$ is estimated from the flux ratio $\phi_i^b / \phi_j^b$ of both the bluer and redder star of the Pickles catalog (\citealt{pickles}) following the annual and overnight modulation statistics (eight first rows). The annual residuals are well above 100 mmag in all bands and between 10-27 mmag overnight. These numbers are to be compared with the estimated MERRA-2 uncertainties, broken up by constituents in (last three rows). The estimated MERRA-2 uncertainty on the ozone field has a negligible impact on the photometry while the impact of the AOD uncertainty is larger than the typical overnight variations, except in the Y4 band. The chromaticity of the aerosols is capture by the Angstrom exponent ($\alpha$) following the same statistics, applied to the 2-D field shown figure \ref{fig:A_AE} appendix \ref{appx2}: it adds a 1 mmag chromatic term in $u$ and $g$ overnight and 3 mmag annually. The LSST will supplement its main broadband telescope with a second telescope equipped with a slitless spectrograph that will be dedicated to real time monitoring of the atmospheric transmission along the survey line of sight. It will thus be able to combine three independent determination of the atmospheric variability: the photometric and spectrometric observations, and reanalysis fields. The residuals reported in table \ref{tab:1}, which provides an estimates of the relative merits of the different MERRA-2 fields, will be useful to derive an atmospheric calibration strategy that combine optimally the various measurements.

\begin{table*}
\caption{\label{tab:1} Millimagnitude chromatic residuals in LSST u, g, r, i, z, y passbands from  observing red and a blue star (Pickles catalog) due to the annual modulation of the atmospheric parameters (four first rows) and overnight variation (next four rows).  The last three rows indicate the propagation of MERRA-2 uncertainties for the ozone, the PWV and the AOD respectively. Uncertainties are below both annual and overnight variations for the ozone and PWV fields, and in between annual and overnight variations for the AOD. The chromaticity of the residuals can be estimated from the difference between the red and blue star residuals with levels above 1 mmag indicated in bold. The angstrom exponent, which carries out the chromatic part of the aerosols is also indicated, although it is estimated from 2-D fields which are less reliable given the abrupt orography surrounding the telescope site.}
\begin{center}
\begin{tabular}{c|ccccccc}
\hline
\hline
\\[-6pt]
[red star - blue star] (mmag.) & u & g & r  & i & z  & y  \\
\hline 
\\[-6pt]
O$_3$ Annual &  \textbf{0 - 1} & \textbf{2 - 1} & \textbf{4 - 5} & 0 - 0 & 0 - 0 & 0 - 0 \\[1pt]
PWV Annual & 0 - 0 & 0 - 0 & \textbf{4 - 3} & \textbf{15 - 17} & \textbf{28 - 26} & \textbf{84 - 95} \\[1pt]
AOD Annual & 137 - 137 & 137 - 137 & 137 - 137 & 137 - 137 & 137 - 137 & 137 - 137 \\[1pt]
$\alpha$ Annual & \textbf{24 - 27} & \textbf{12 - 15} & \textbf{6 - 7} & 3 - 3 & 1 - 1 & 0 - 0 \\[1pt]
\hline
\\[-6pt]
O$_3$ Nightly & 0 - 0 & 0 - 0 & 1 - 1 & 0 - 0 & 0 - 0 & 0 - 0 \\[1pt]
PWV Nightly & 0 - 0 & 0 - 0 & 1 - 1 & \textbf{4 - 5} & \textbf{8 - 7} & \textbf{23 - 26} \\[1pt] 
AOD Nightly & 10 - 10 & 10 - 10 & 10 - 10 & 10 - 10 & 10 - 10 & 10 - 10 \\[1pt]
$\alpha$ Nightly & \textbf{8 - 9} & \textbf{4 - 5} & 2 - 2 & 1 - 1 & 0 - 0 & 0 - 0 \\[1pt]
\hline
\hline
\\ [-6pt]
0$_3$ (8 DU uncert.)  & 0 - 0 & 0 - 0 & 1 - 1 & 0 - 0 & 0 - 0 & 0 - 0 \\[1pt]
PWV (1 mm uncert.) & 0 - 0 & 0 - 0 & 1 - 1 & 3 - 3 & 4 - 4 & \textbf{13 - 15} \\[1pt]
AOD (0.02 uncert.) & 22 - 22 & 22 - 22 & 22 - 22 & 22 - 22 & 22 - 22 & 22 - 22 \\[2pt]
\hline
\end{tabular}
\tablefoot{Bold numbers indicate chromaticity above the mmag level (Pickles redder star - bluer star $\ge$ 1 mmag).
 }
\end{center}
\end{table*}

\begin{acknowledgements}
This material is based upon work supported in part by the National Science Foundation through Cooperative Agreement 1258333 managed by the Association of Universities for Research in Astronomy (AURA), and the Department of Energy under Contract No. DE-AC02-76SF00515 with the SLAC National Accelerator Laboratory. Additional LSST funding comes from private donations, grants to universities, and in-kind support from LSSTC Institutional Members. The MERRA-2 data set source are provided by NASA/GSFC, Greenbelt, MD, USA, NASA Goddard Earth Sciences Data and Information Services Center (GES DISC). The authors thanks the LSST internal reviewers Eli Rykoff and Peter Yoachim for their contribution to the peer review of this work.
\end{acknowledgements}

\bibliographystyle{aa}
\bibliography{biblio}

\appendix
\appendixpage
\addappheadtotoc

\section{MERRA-2 parameters distributions and profiles above Mauna Kea and CTIO}
\label{appx1}

\begin{figure}
\centering
\includegraphics[width=0.25\linewidth]{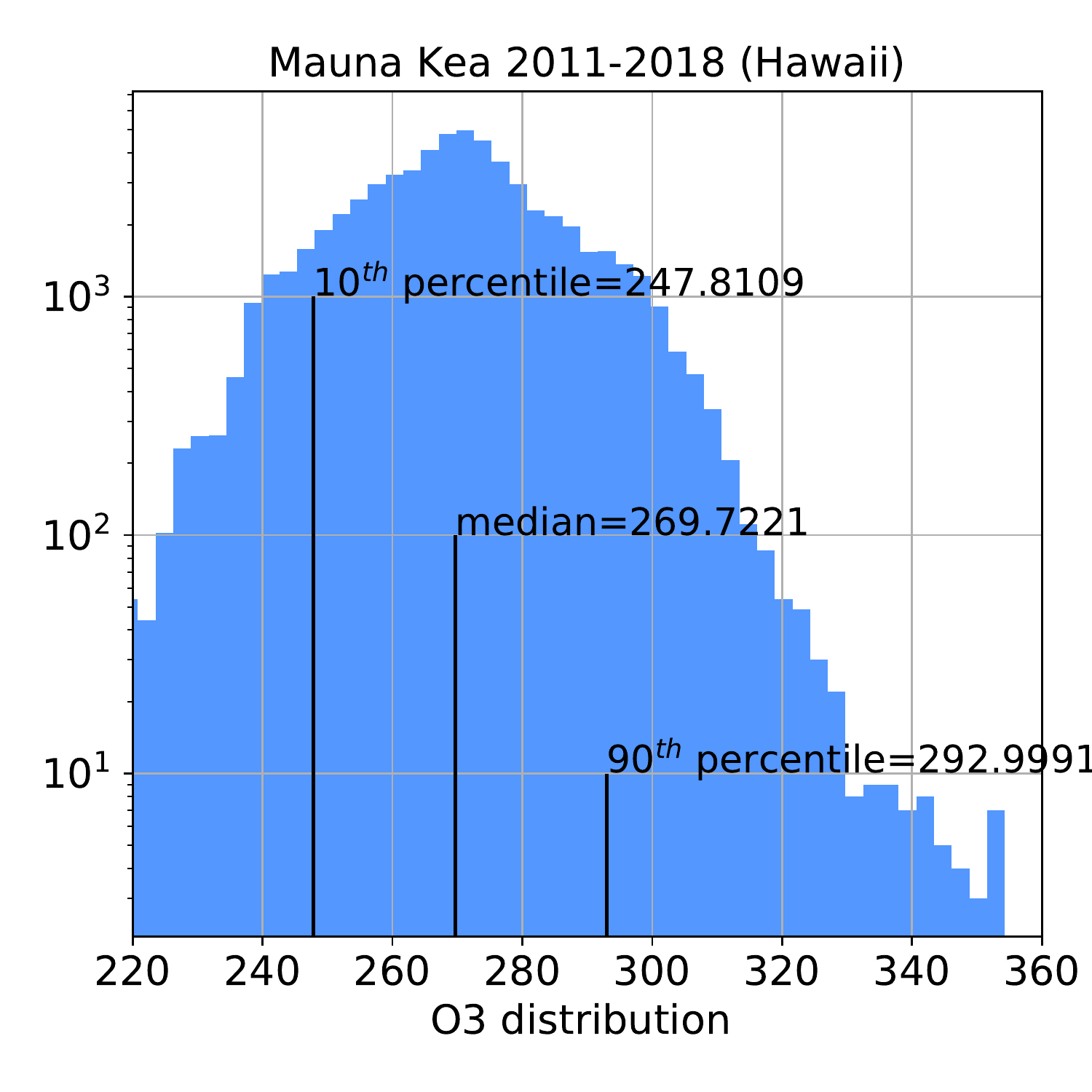}
\includegraphics[width=0.25\linewidth]{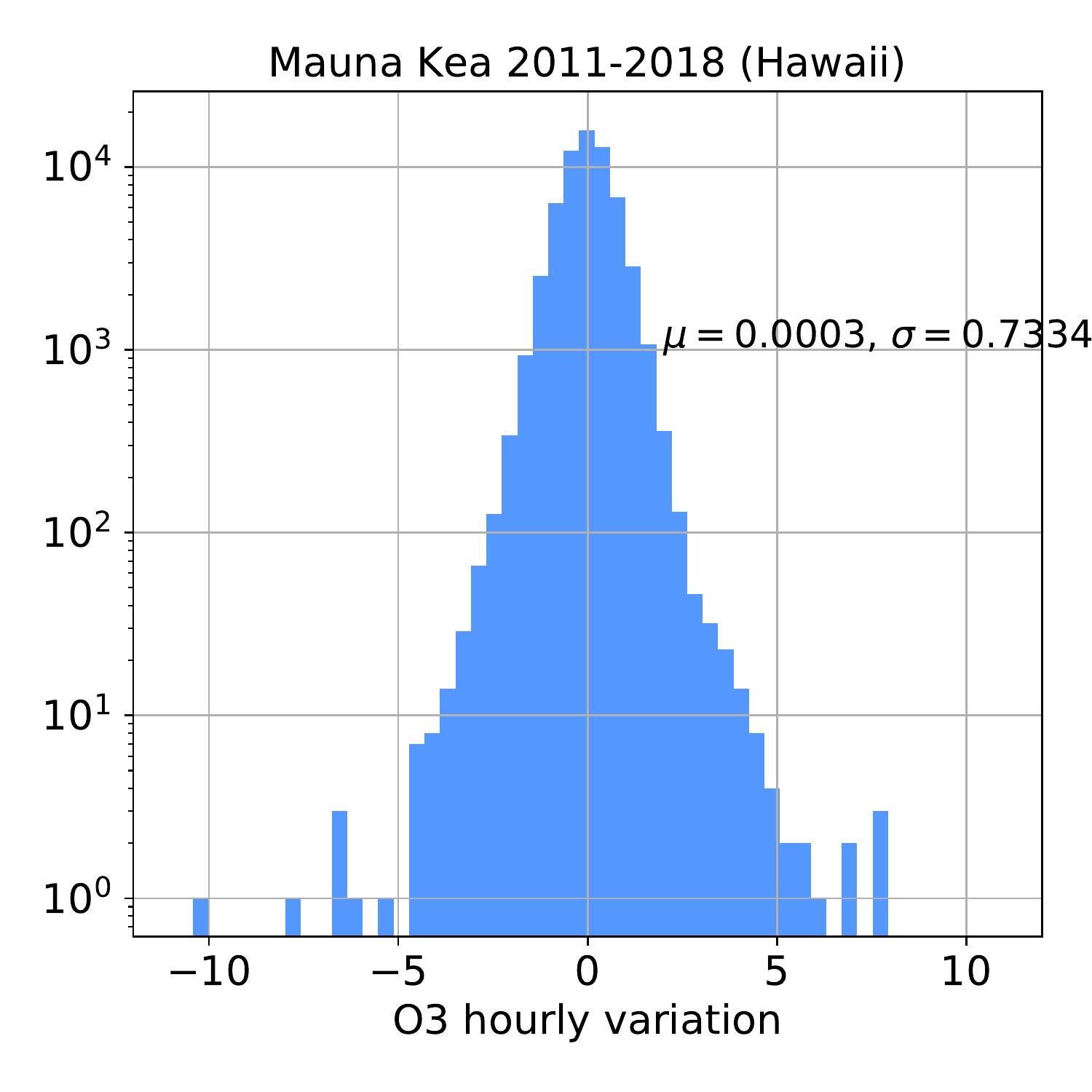}
\includegraphics[width=0.25\linewidth]{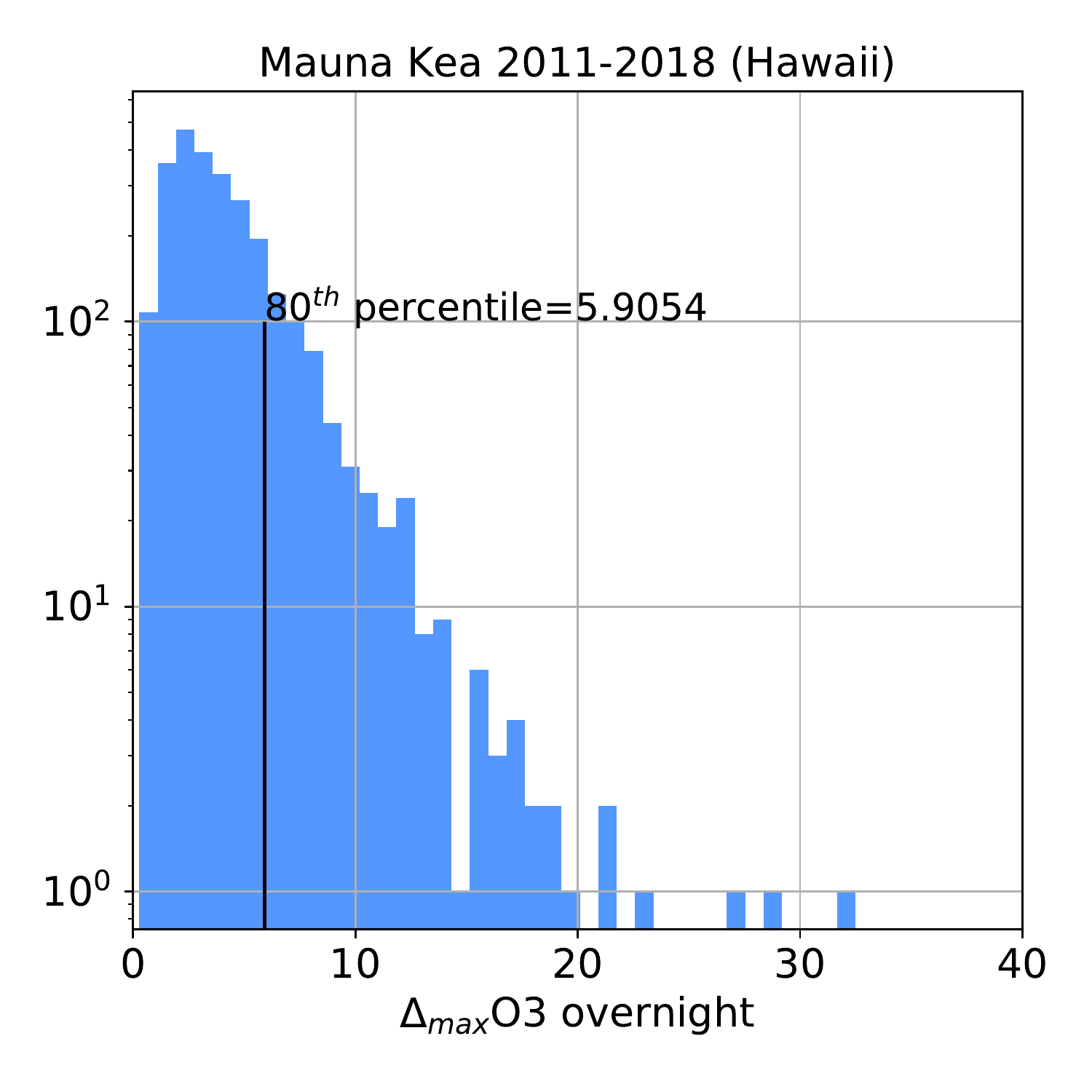}
\includegraphics[width=0.25\linewidth]{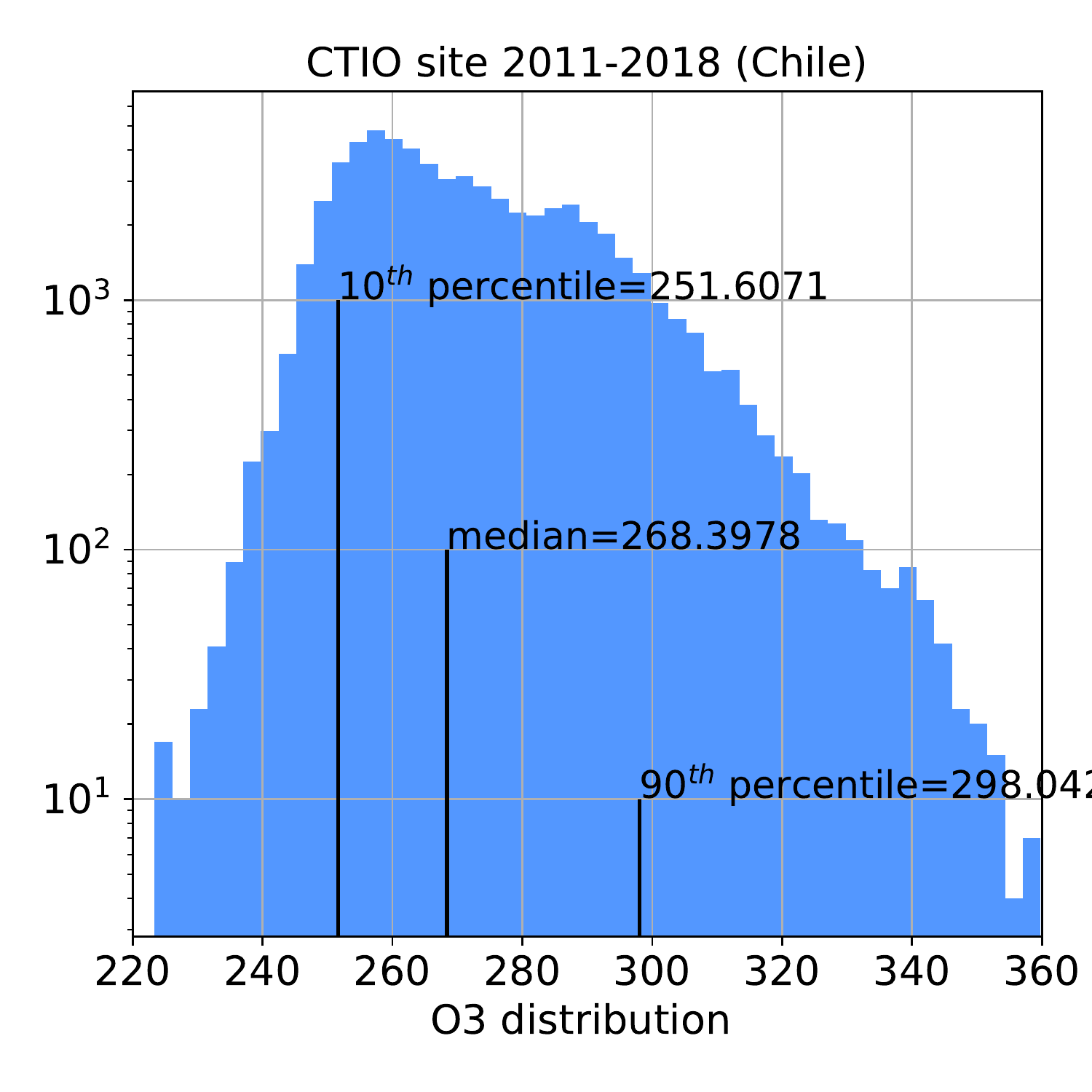}
\includegraphics[width=0.25\linewidth]{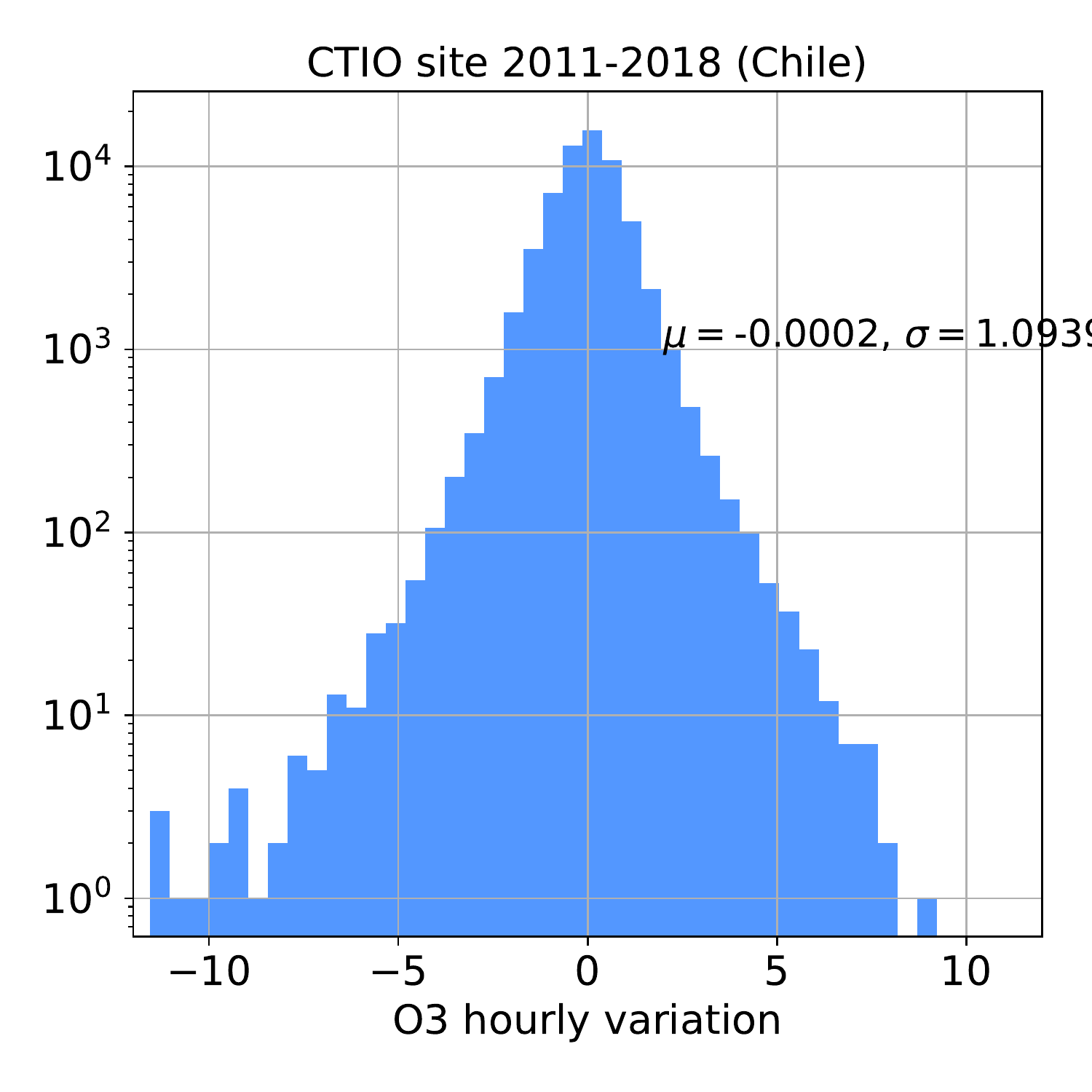}
\includegraphics[width=0.25\linewidth]{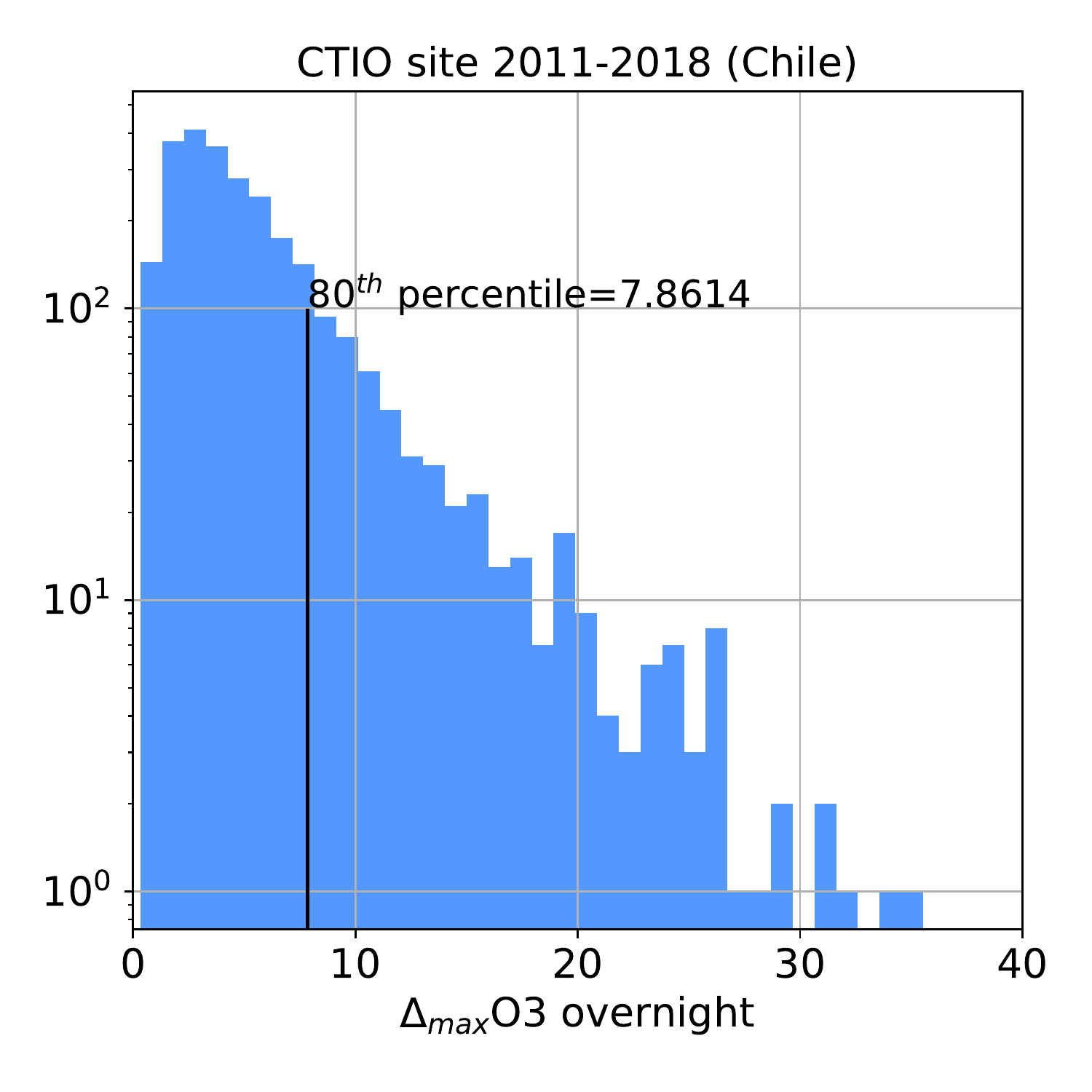}
\caption{Ozone distribution (left), hourly variation distribution (center) and maximum overnight variation distribution (right) above Mauna Kea (upper panels) and CTIO (lower panels).
 \label{fig:Ao3}}
\end{figure}

\begin{figure}
\centering
\includegraphics[width=0.25\linewidth]{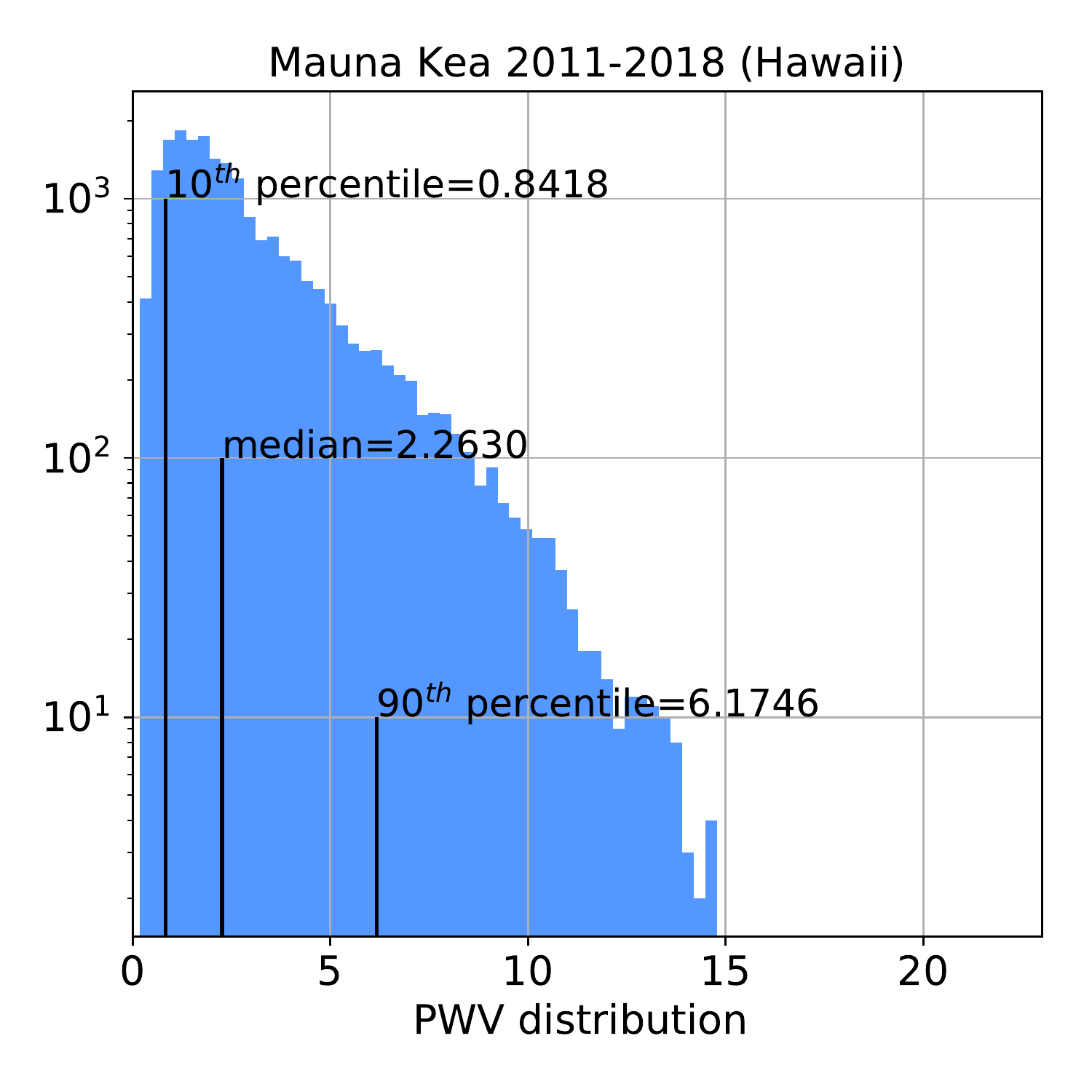}
\includegraphics[width=0.25\linewidth]{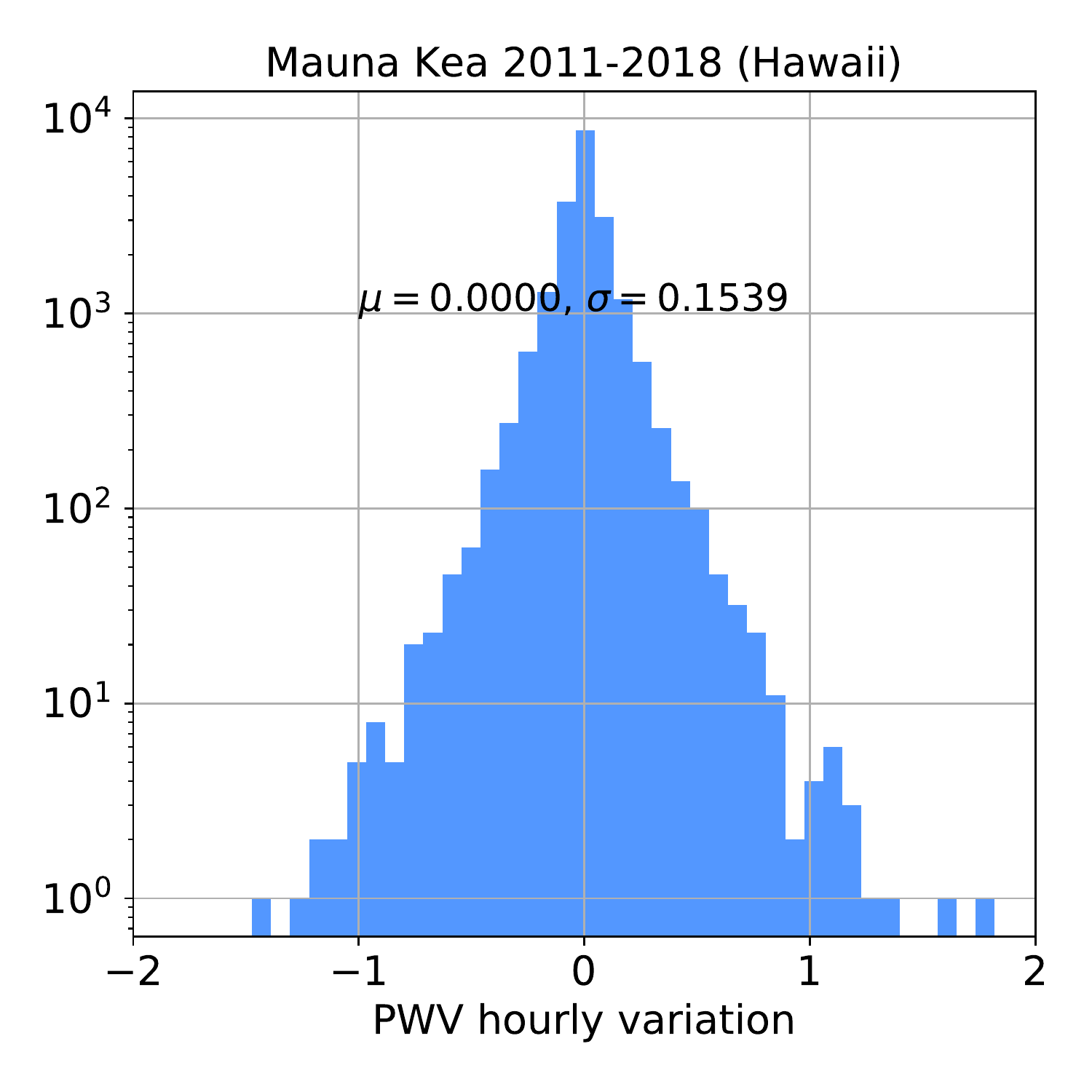}
\includegraphics[width=0.25\linewidth]{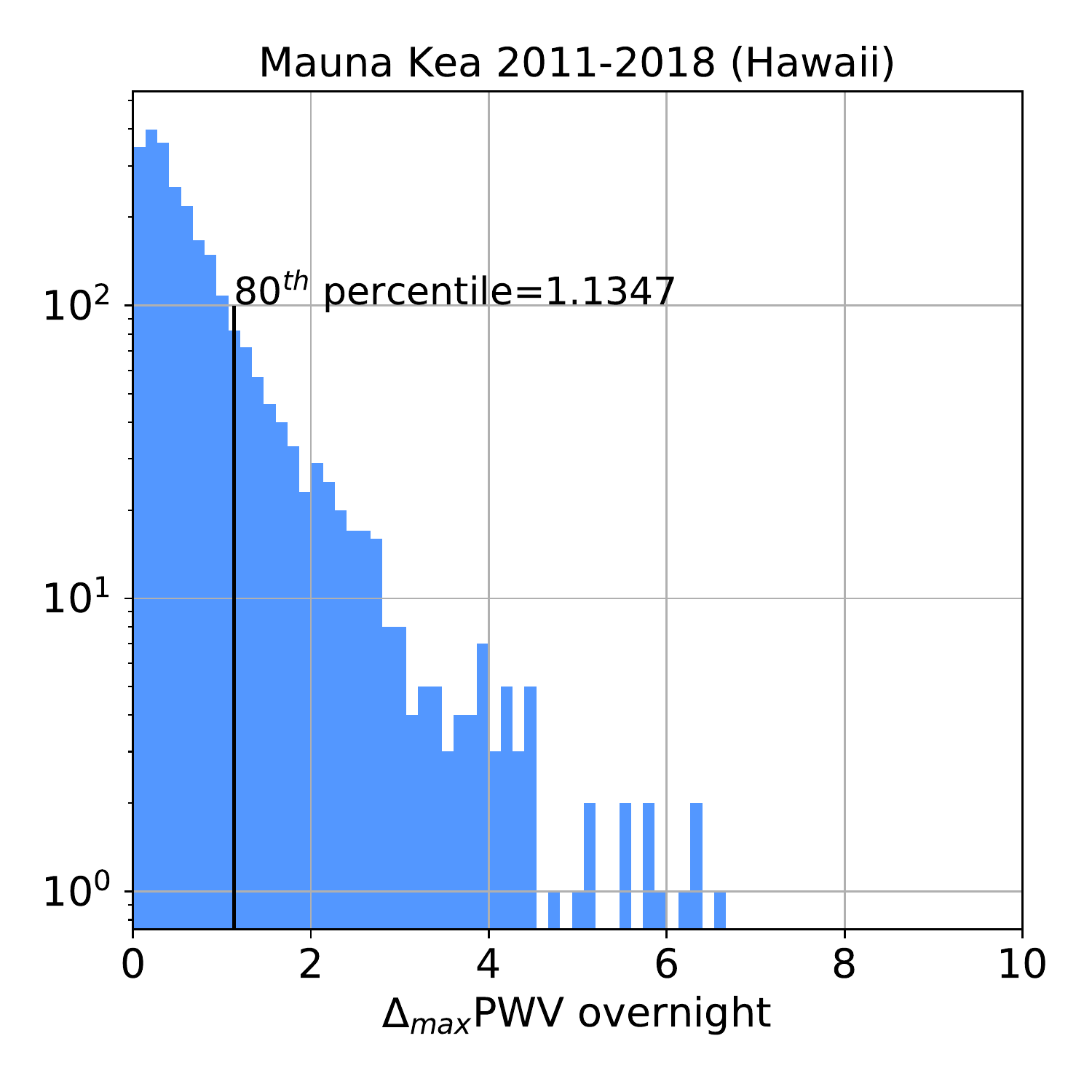}
\includegraphics[width=0.25\linewidth]{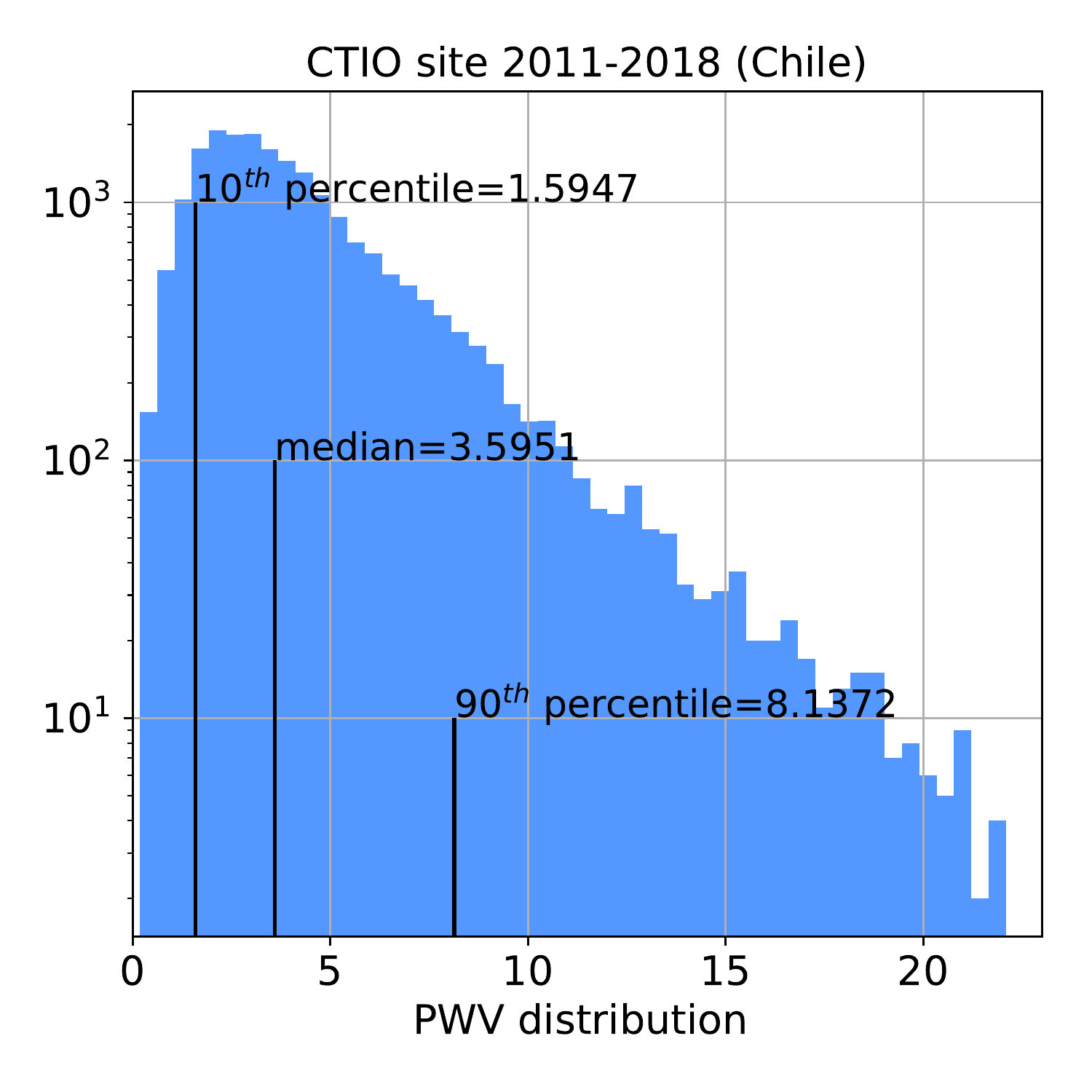}
\includegraphics[width=0.25\linewidth]{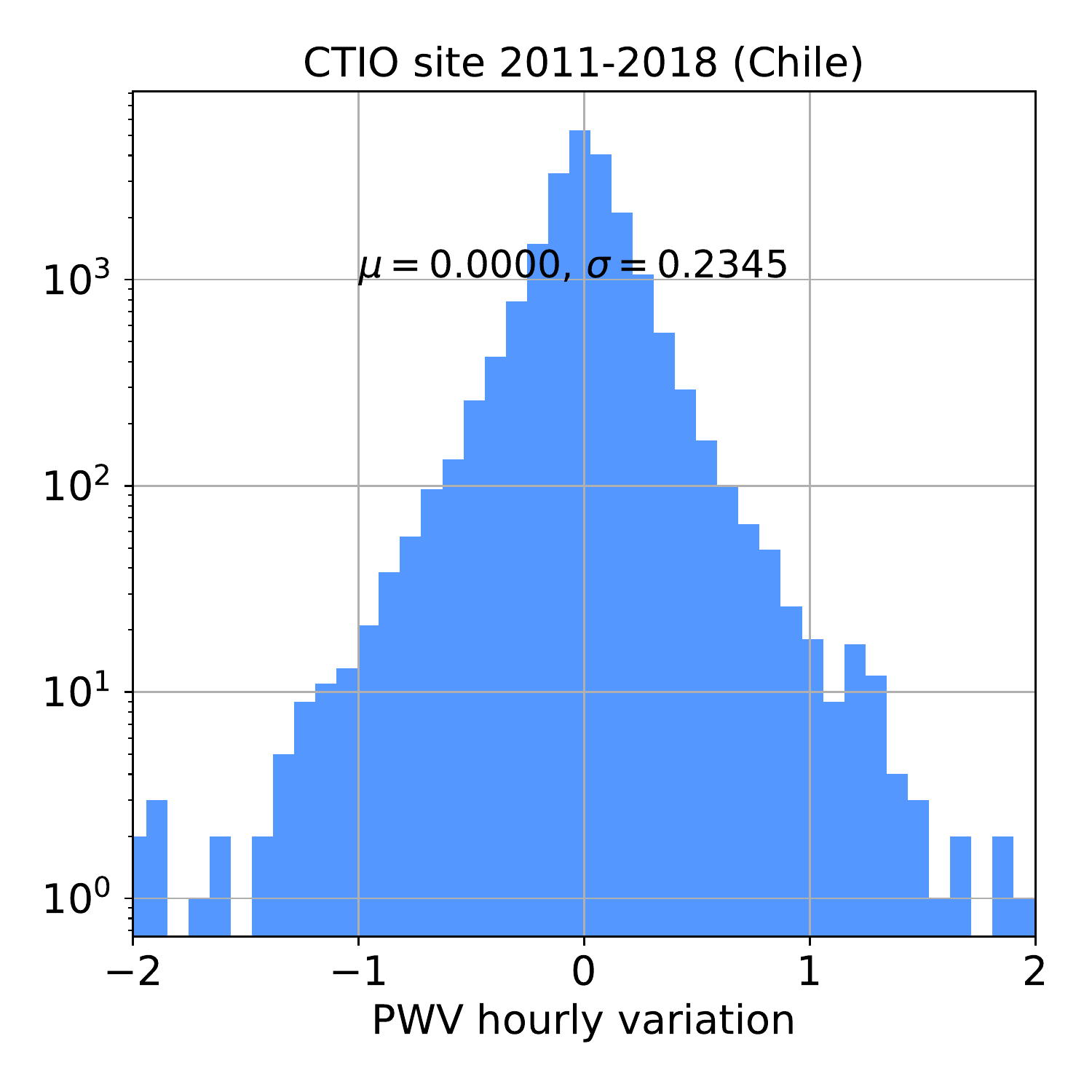}
\includegraphics[width=0.25\linewidth]{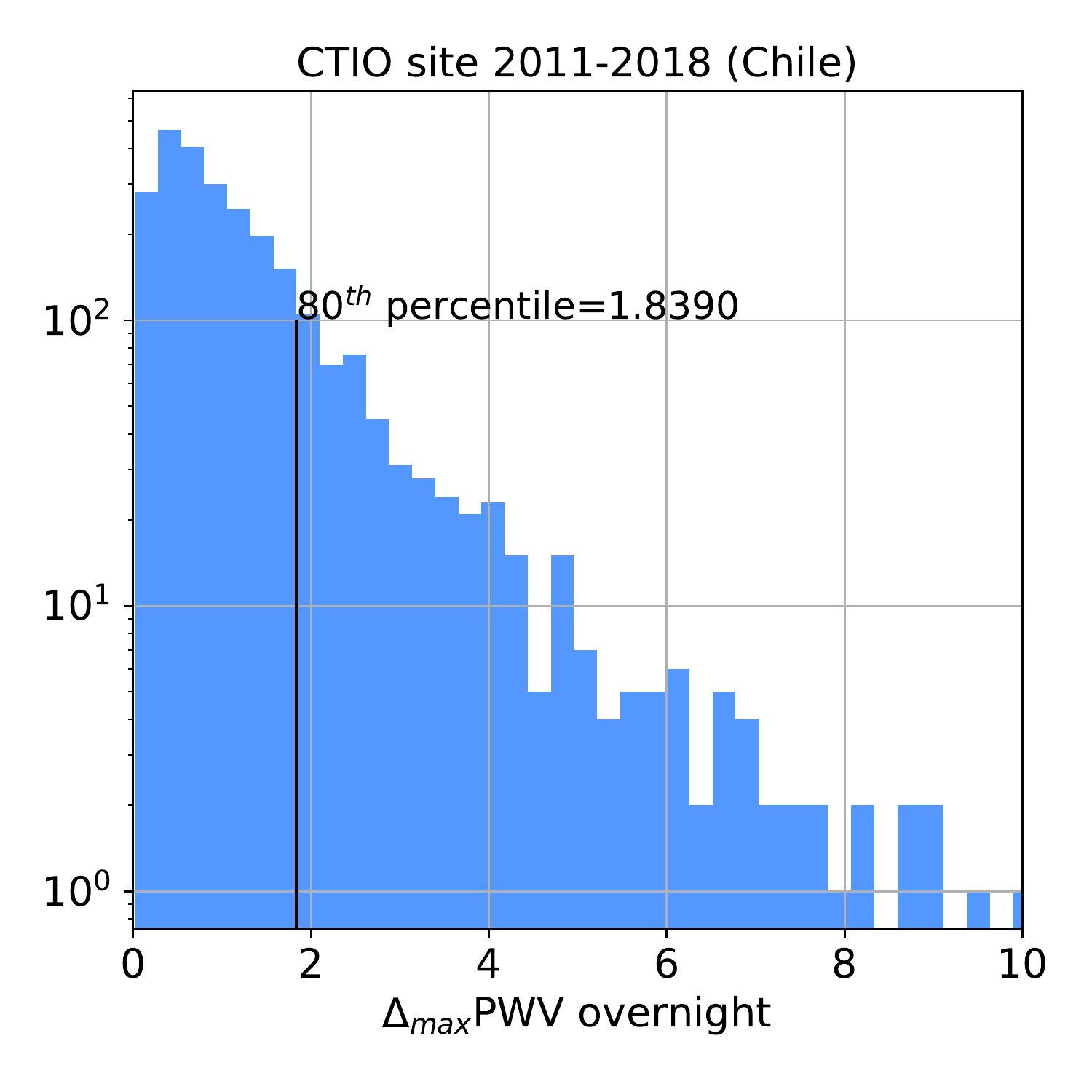}
\caption{Distributions of millimeter of precipitable water vapor (left), hourly variation (center) and maximum overnight variation (right) above Mauna Kea (upper panels) and CTIO (lower panels) recorder by MERRA-2 between 2011 and 2018. The quality of the Mauna Kea site is significantly better irrespective of the statistics being considered.
 \label{fig:Apwv}}
\end{figure}

\begin{figure}
\centering
\includegraphics[width=0.25\linewidth]{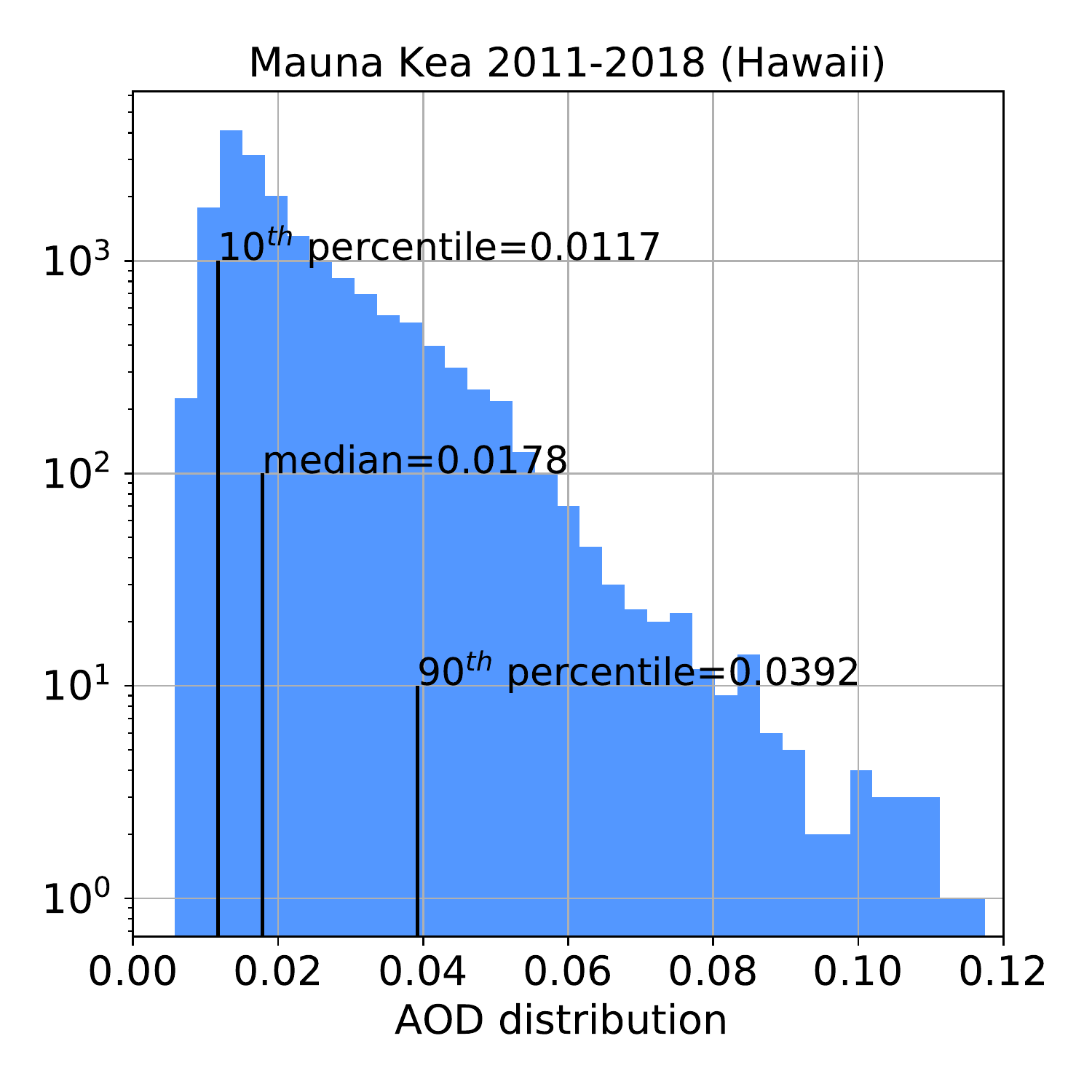}
\includegraphics[width=0.25\linewidth]{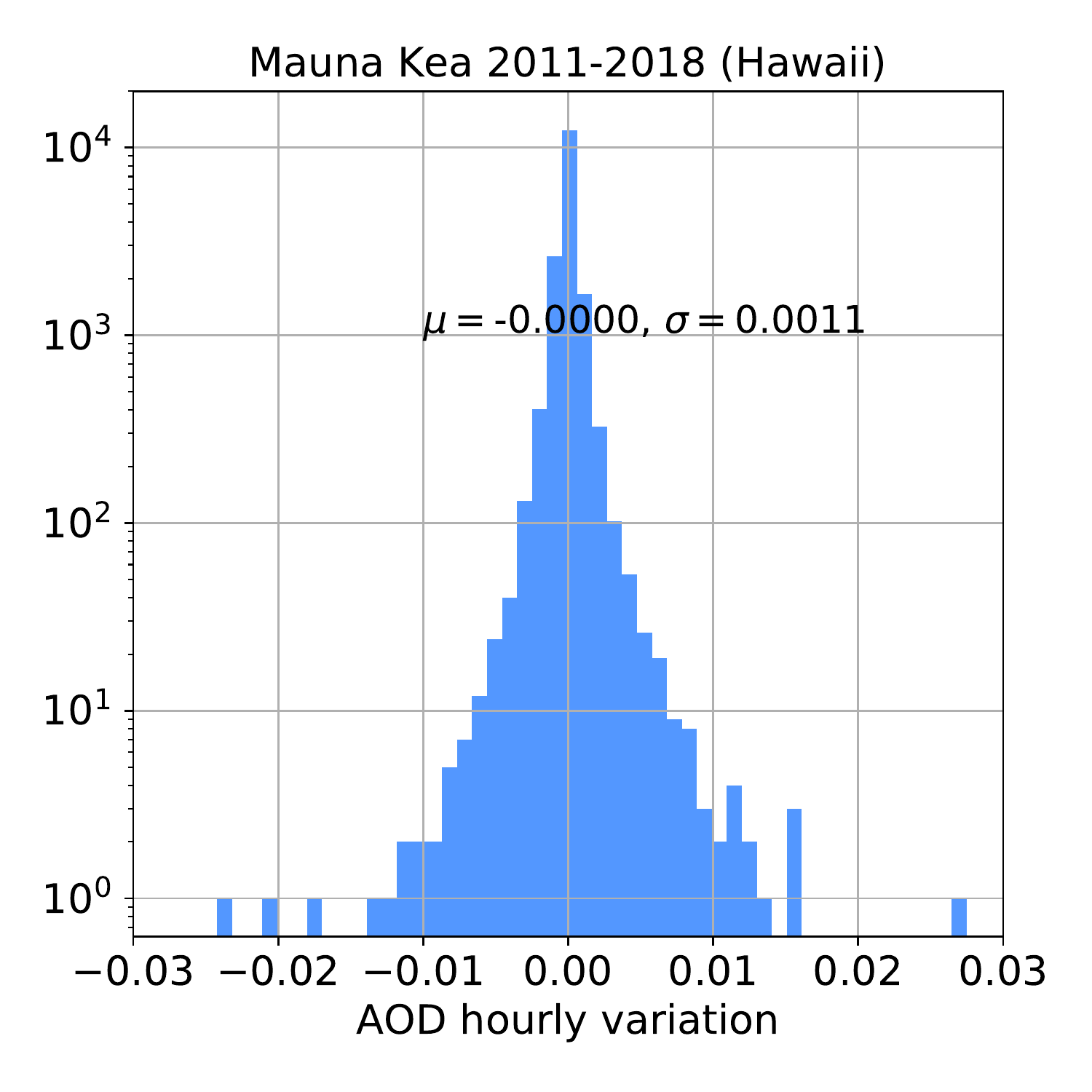}
\includegraphics[width=0.25\linewidth]{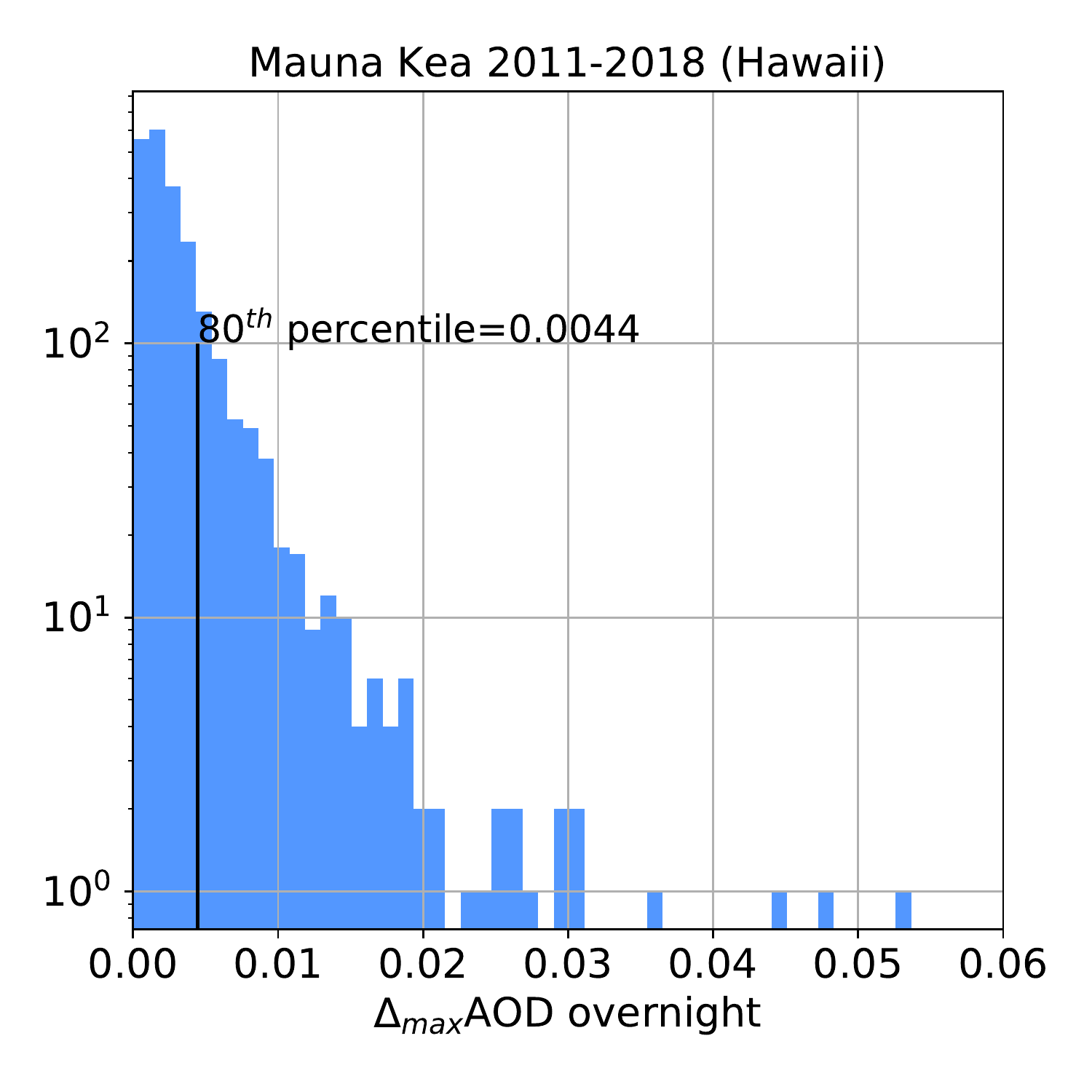}
\includegraphics[width=0.25\linewidth]{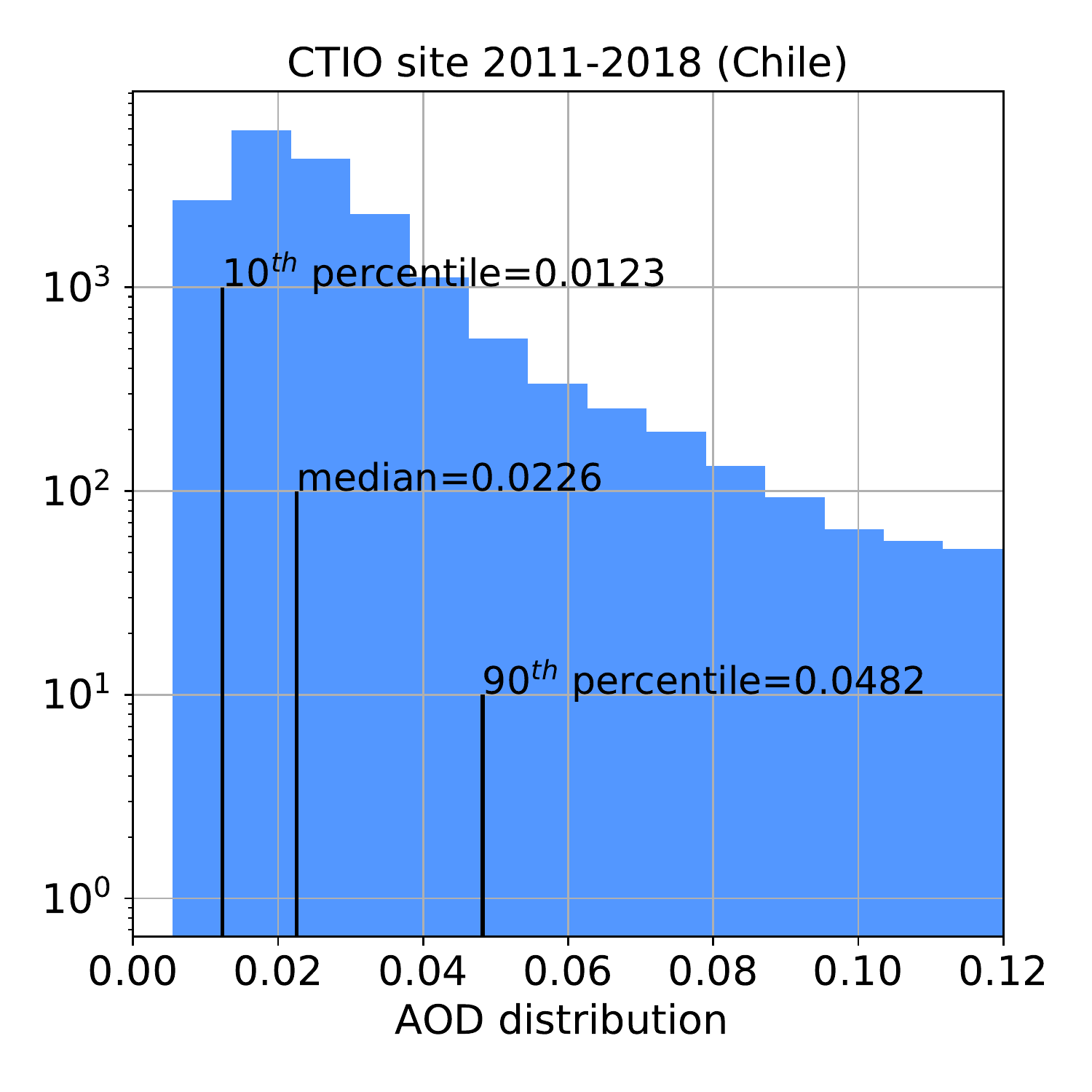}
\includegraphics[width=0.25\linewidth]{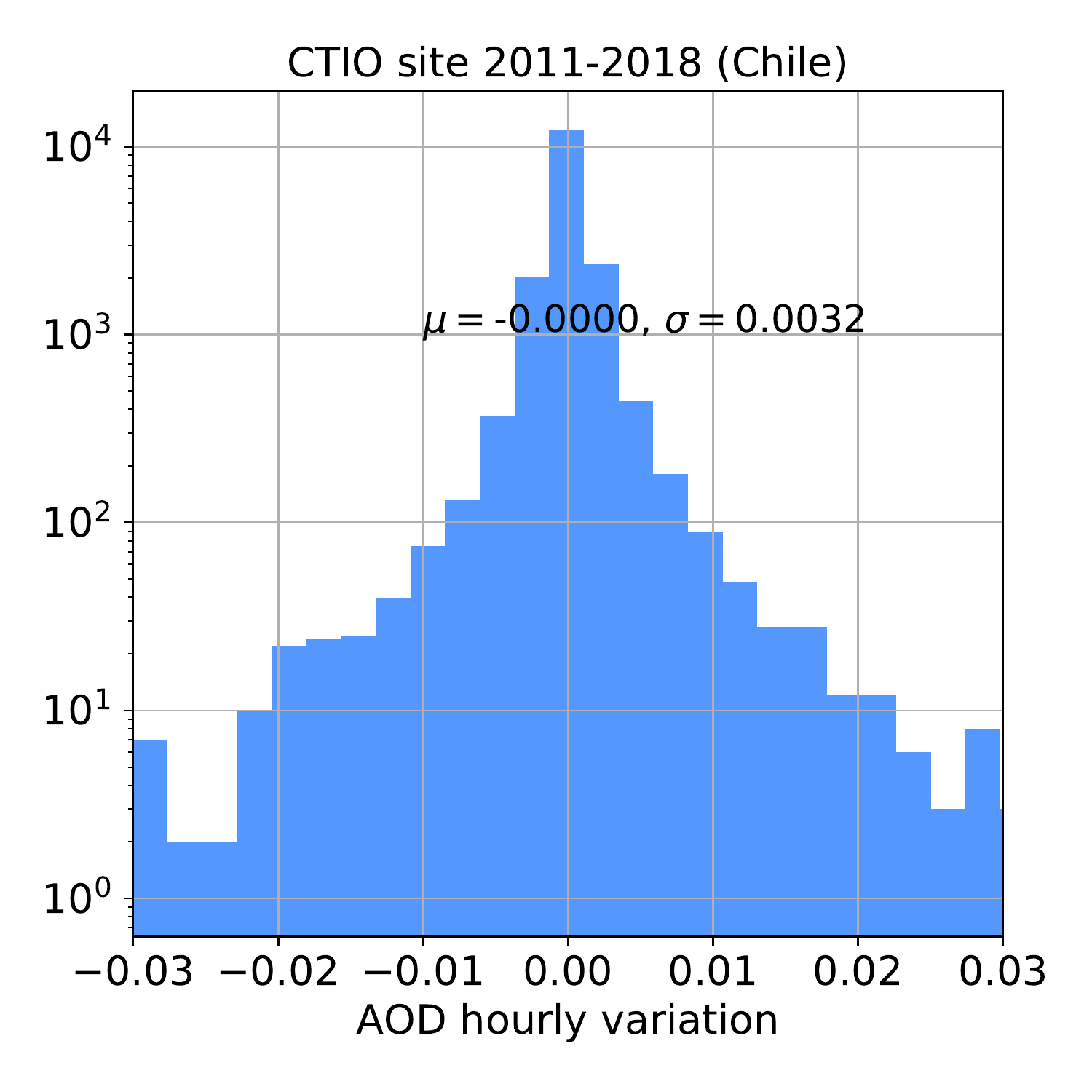}
\includegraphics[width=0.25\linewidth]{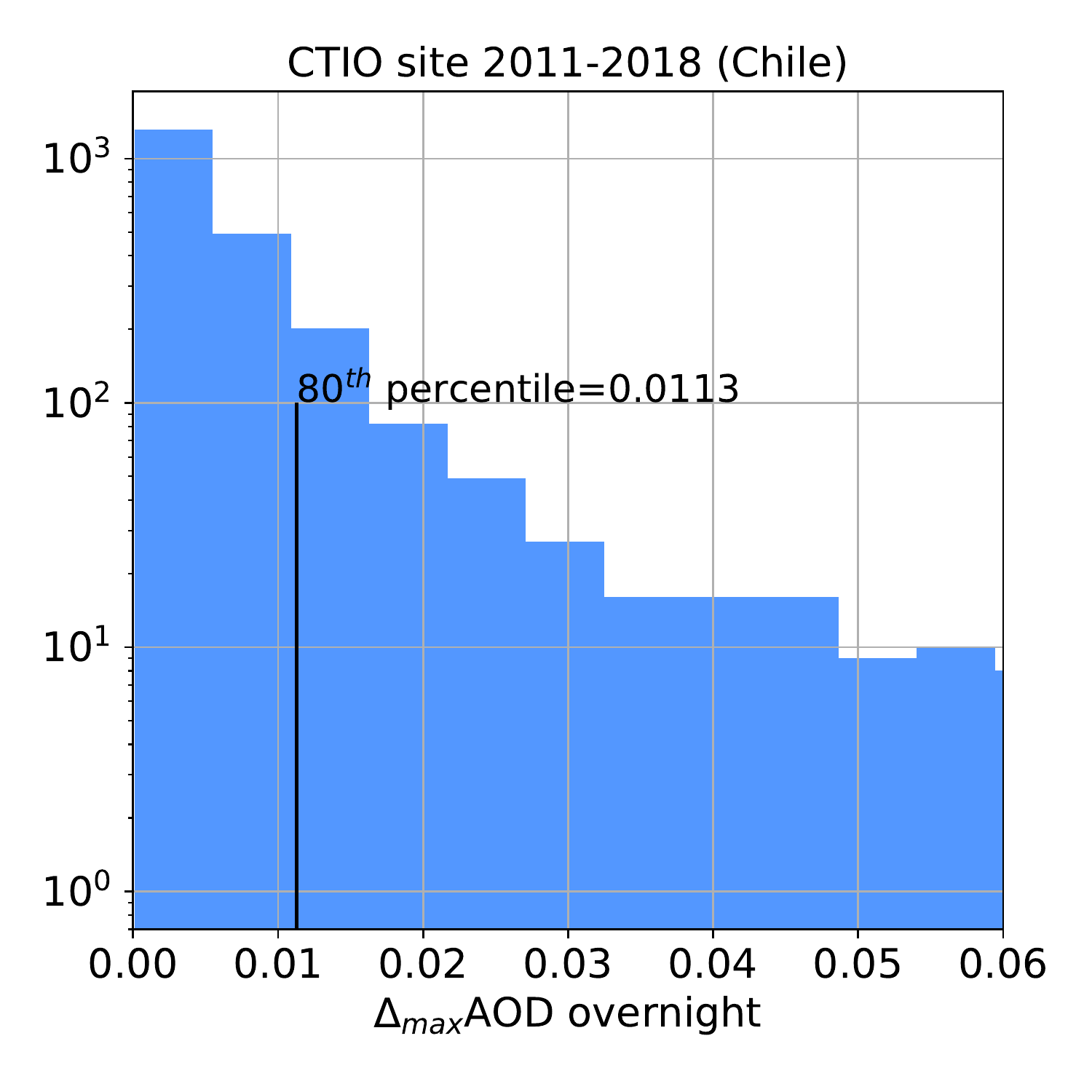}
\caption{Aerosol optical depth distribution (left), hourly variation distribution (center) and maximum overnight variation distribution (right) above Mauna Kea (upper panels) and CTIO (lower panels).
 \label{fig:Aaod}}
\end{figure}

\begin{figure}
\centering
\includegraphics[width=0.4\linewidth]{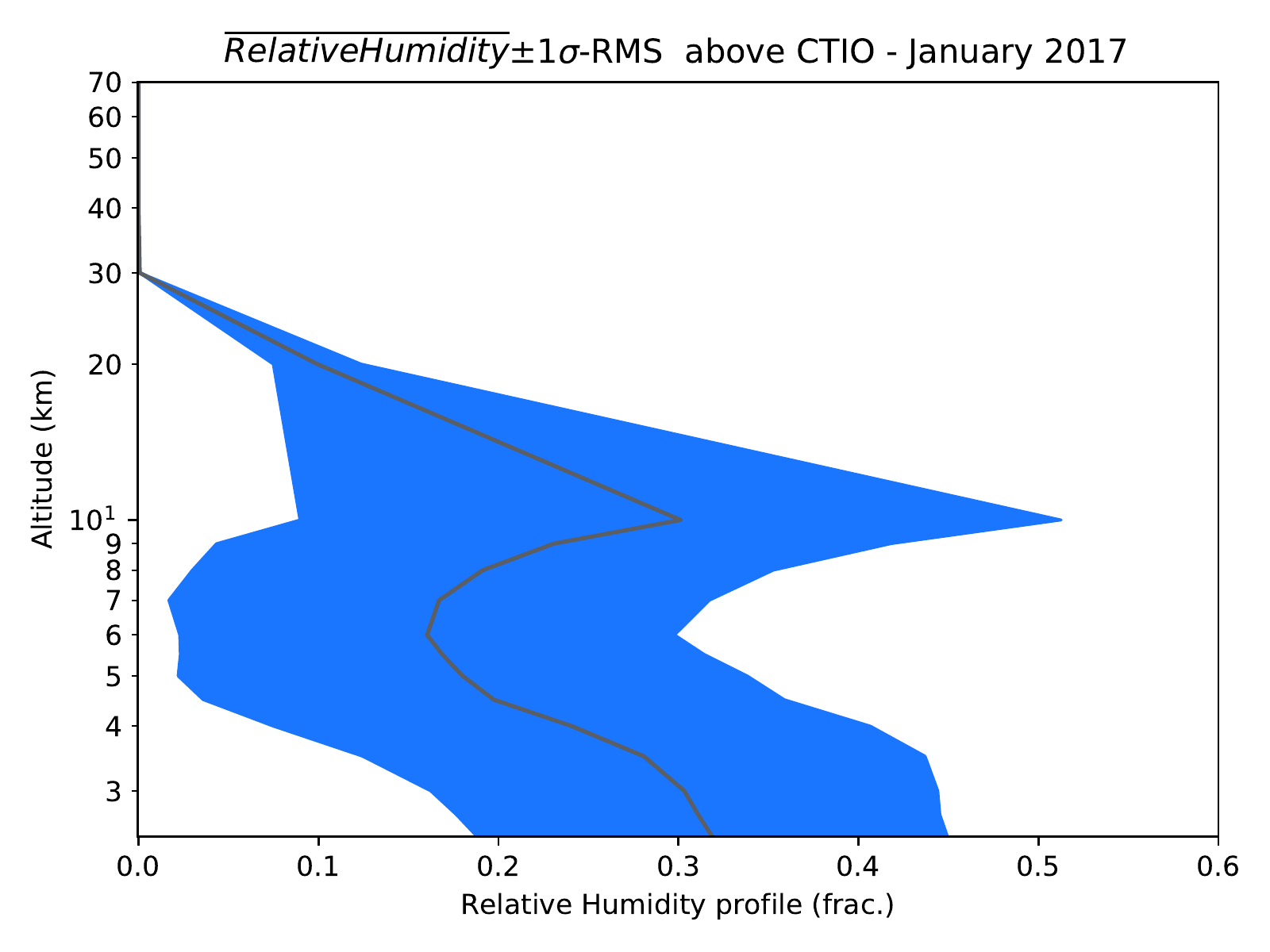}
\includegraphics[width=0.4\linewidth]{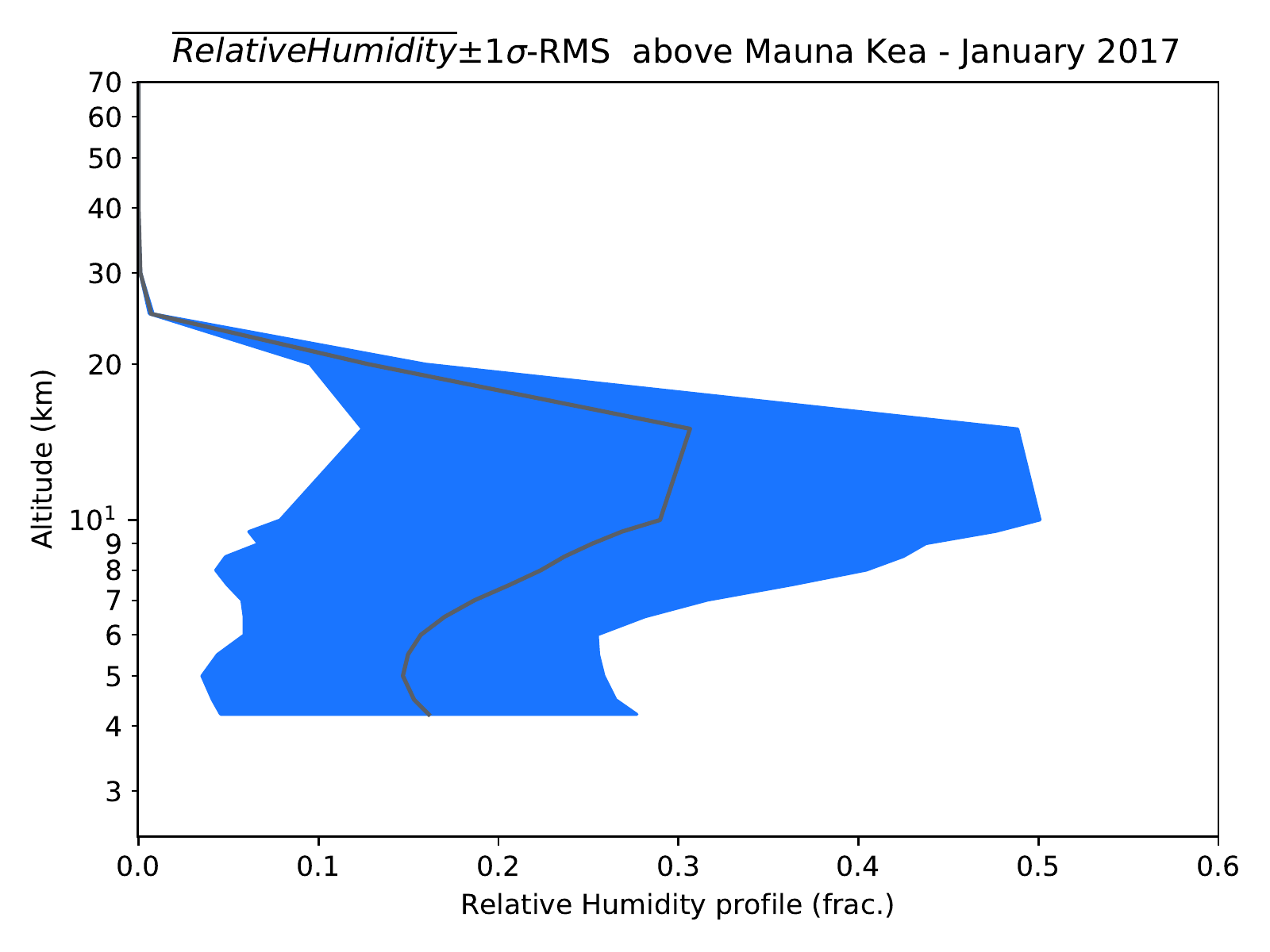}
\includegraphics[width=0.4\linewidth]{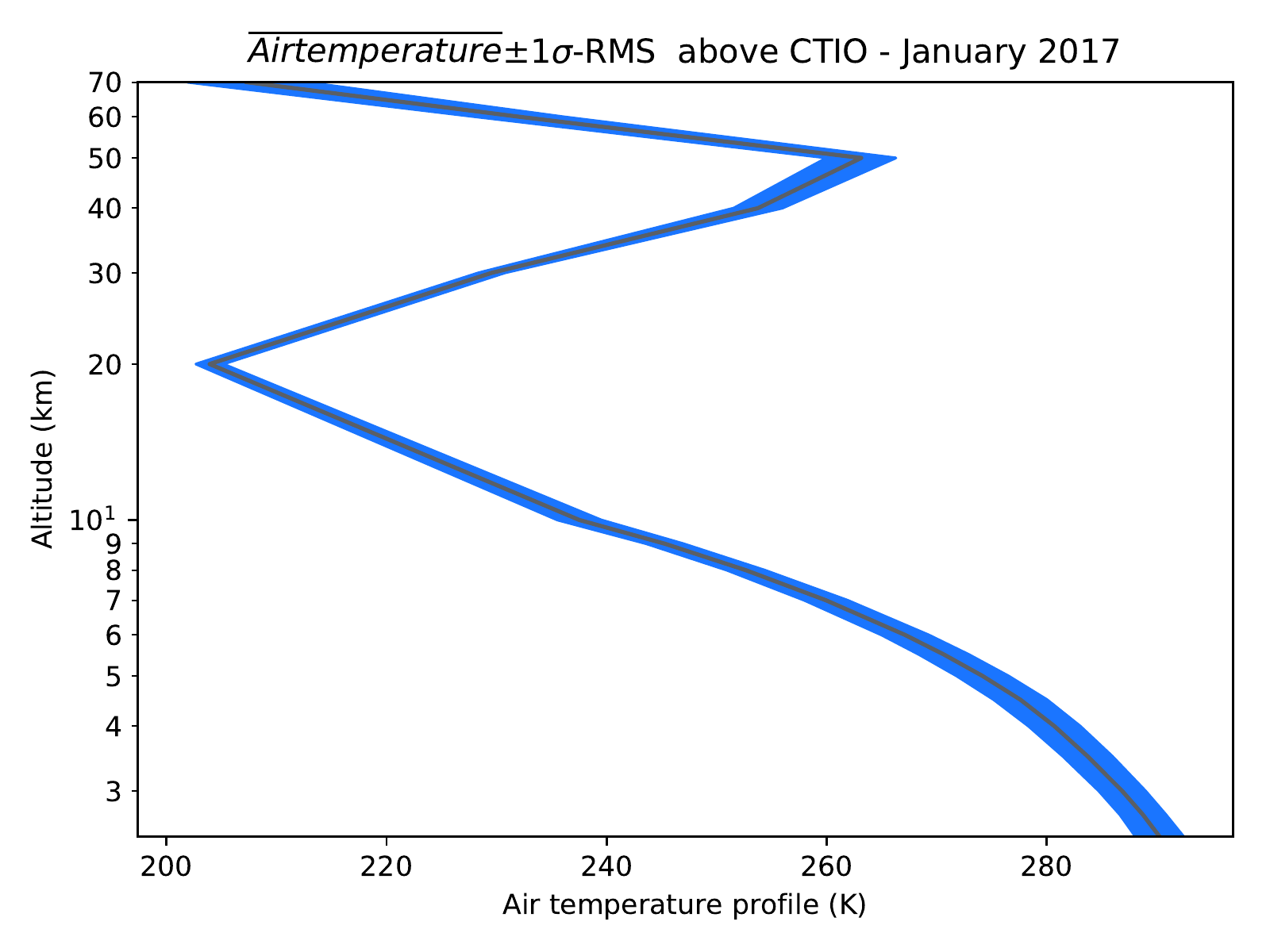}
\includegraphics[width=0.4\linewidth]{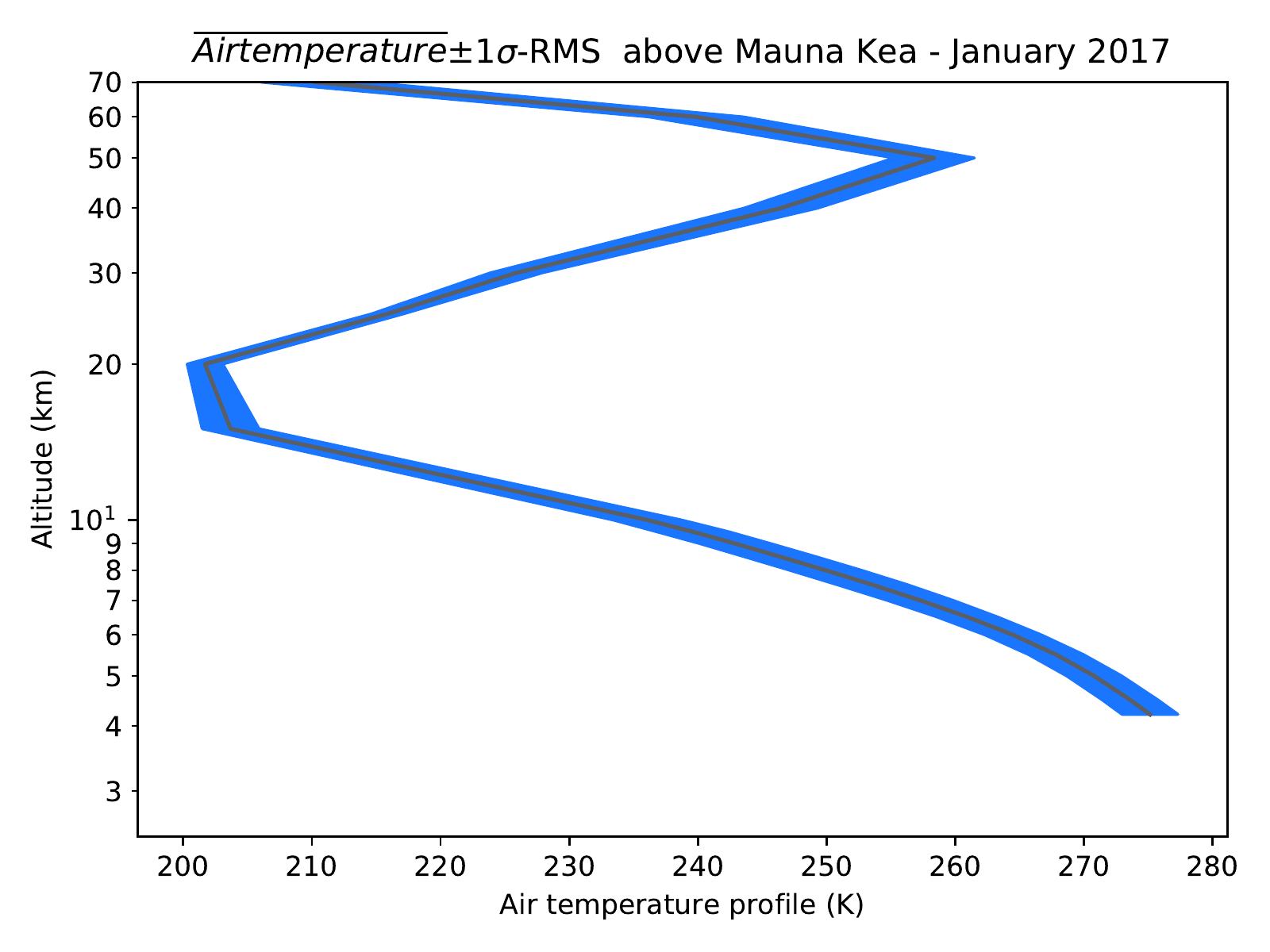}
\caption{Top panels: Tri-hourly relative humidity profiles above CTIO (left) and Mauna Kea sites (right) during January 2017. Relative humidity is the ratio of the partial pressure of water vapor to the equilibrium vapor pressure, which changes with temperature (bottom panels): as a result, a given water vapor amount results in higher relative humidity in cool air than warm air.
 \label{fig:10}}
\end{figure}

\begin{figure}
\centering
\includegraphics[width=0.4\linewidth]{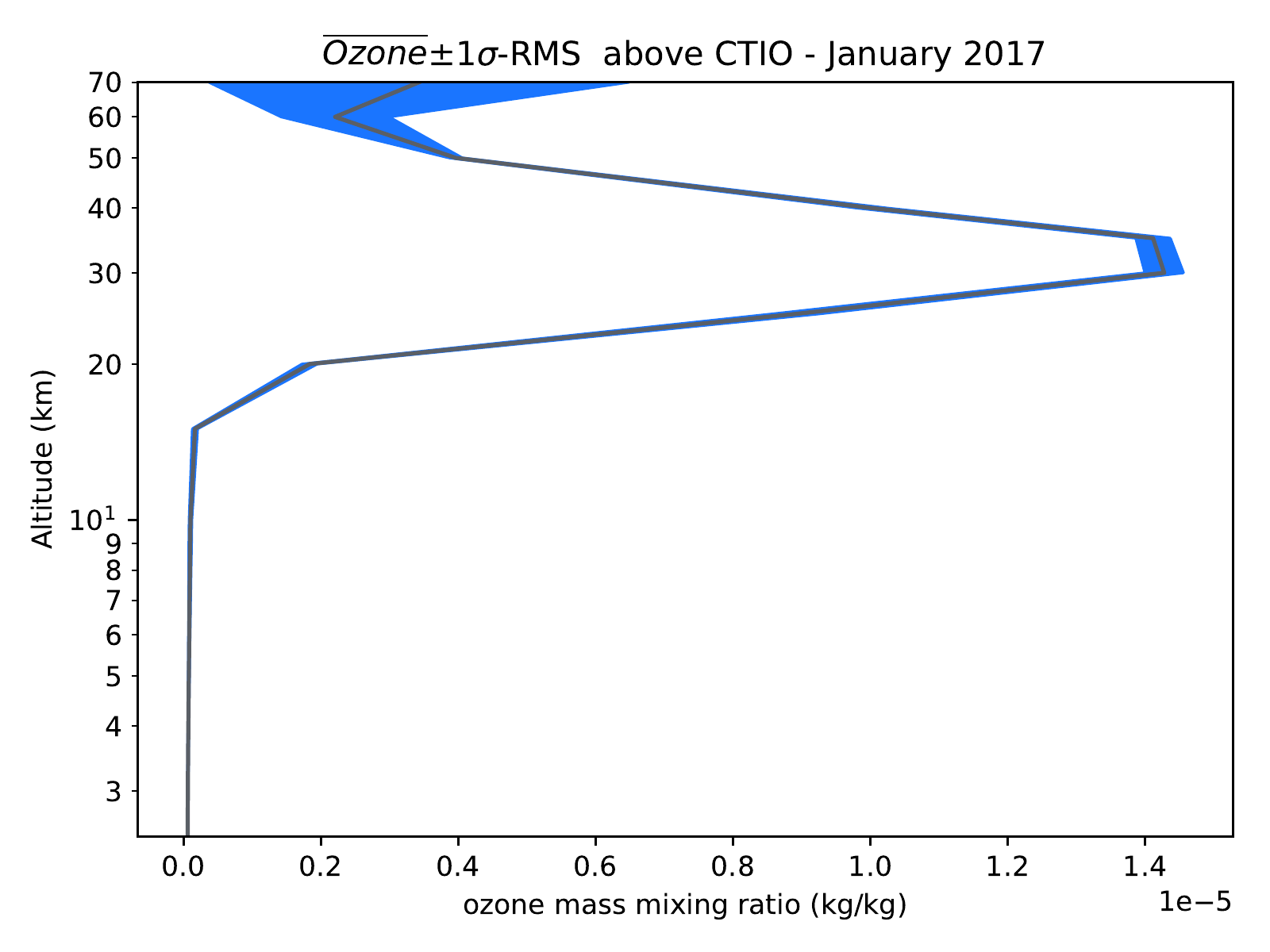}
\includegraphics[width=0.4\linewidth]{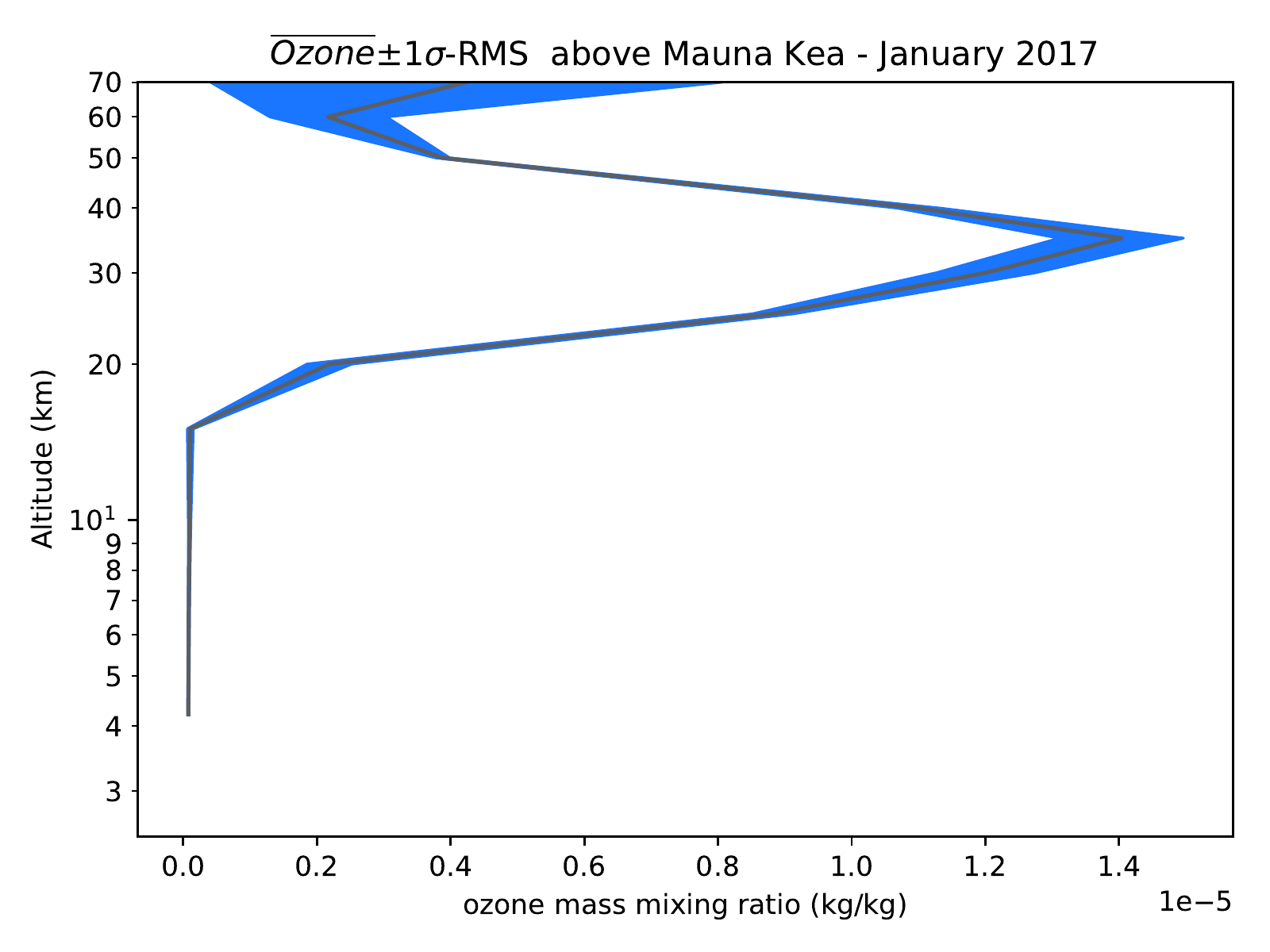}
\caption{Hourly ozone profiles above CTIO (left) and Mauna Kea sites (right) during January 2017.
 \label{fig:A_o3}}
\end{figure}

\section{Aerosol angstrom exponent 2-D fields}
\label{appx2}

\begin{figure}
\centering
\includegraphics[width=0.4\linewidth]{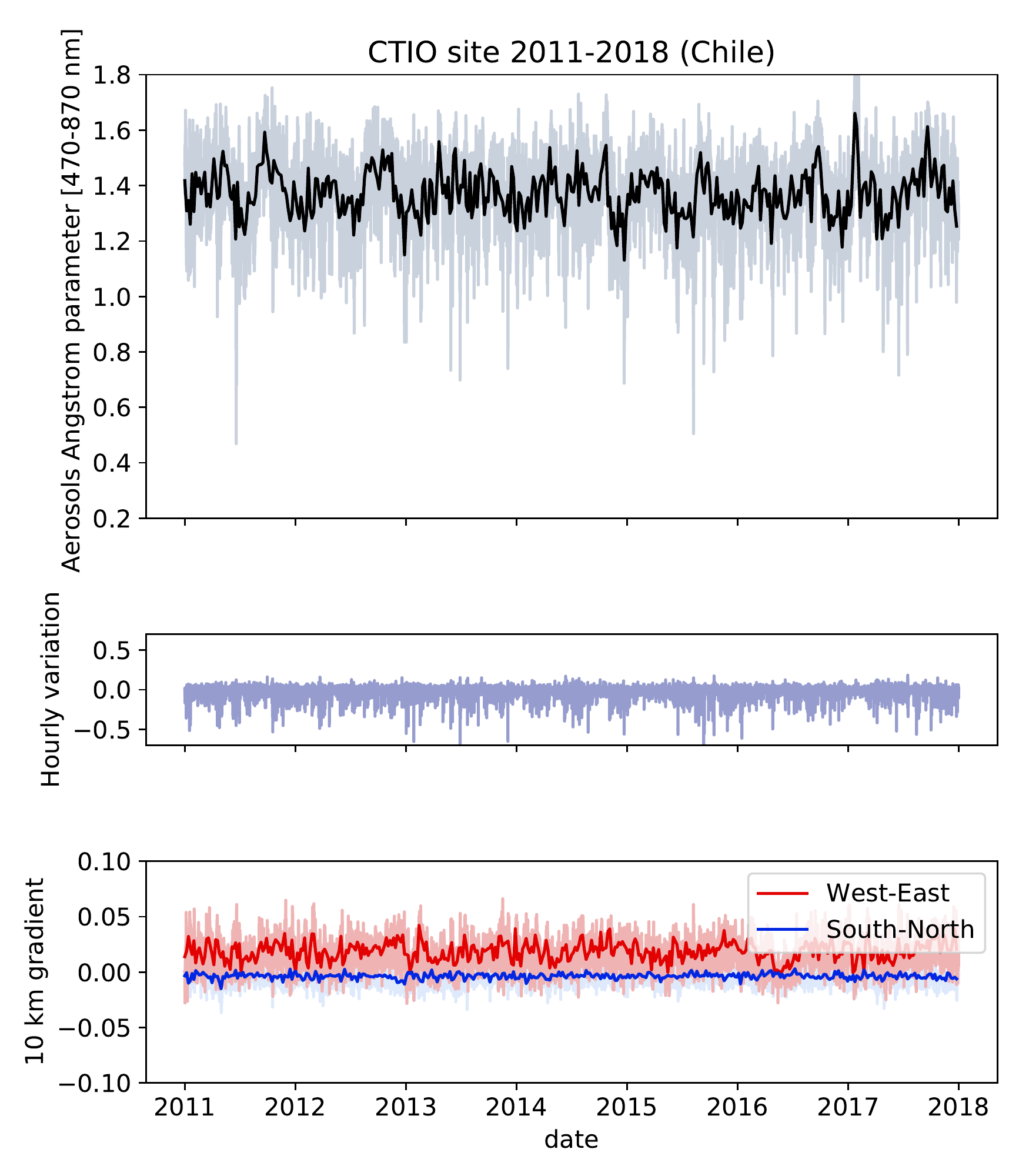}
\includegraphics[width=0.4\linewidth]{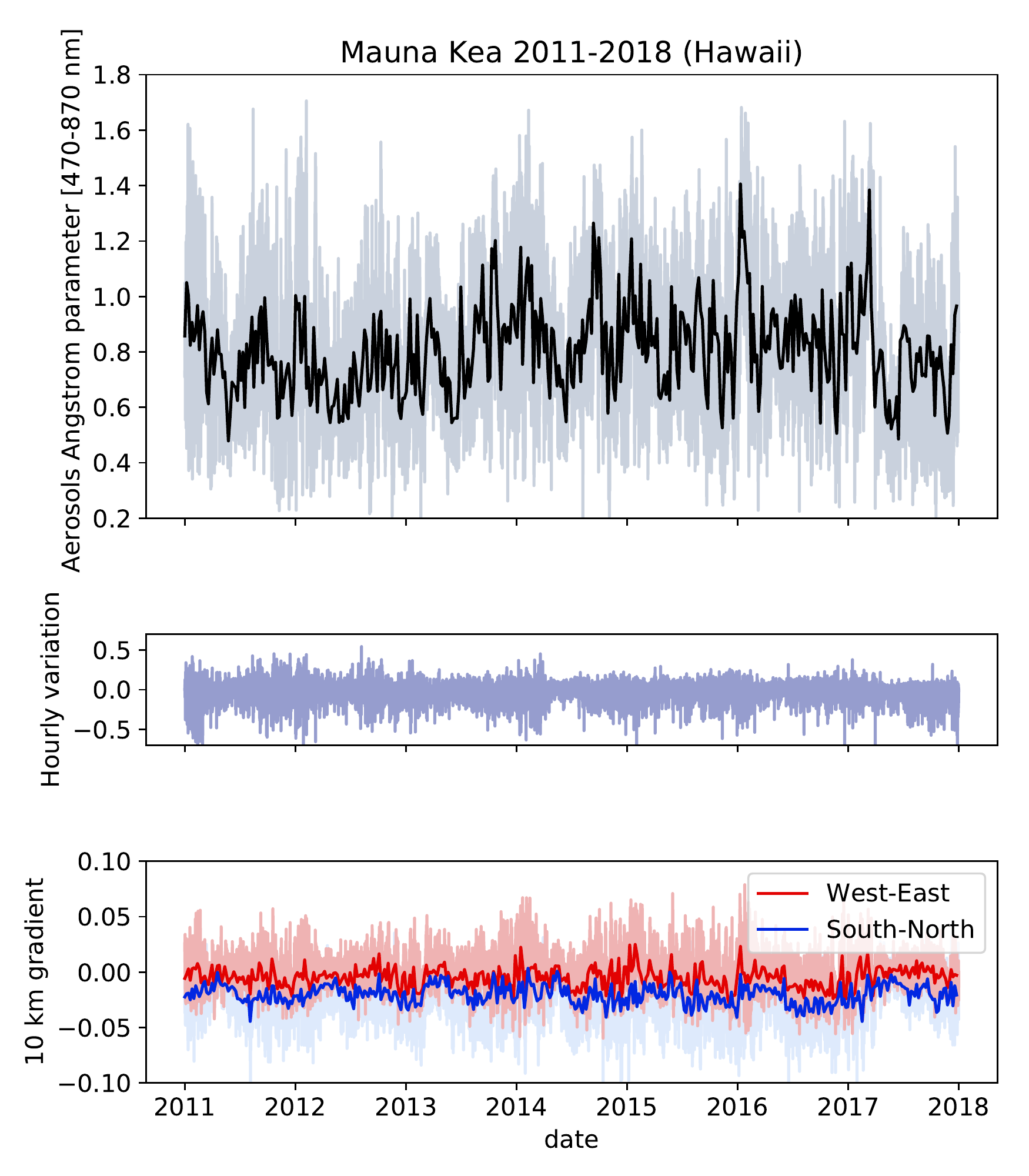}
\caption{Top panels: Aerosol total angstrom exponent between 2011 and 2018 at CTIO (left) and Mauna Kea (right) latitude-longitude coordinates obtained from MERRA-2 2-D tables. The angstrom exponent median value at the CTIO is 1.38 with variability 0.31 (10 to 90 percentile). The overnight modulation is 0.11 (|$\Delta_{max}^{overnight}$| 80 percentile). The mean South-North gradient over the period is -0.004 $\pm$ 0.01/10km, and a West-East gradient at 0.02 $\pm$ 0.01/10km (bottom left panel). 
The angstrom exponent median value at the Mauna Kea summit is 0.78 with variability 0.59 (10 to 90 percentile). The overnight modulation is  0.24 (|$\Delta_{max}^{overnight}$| 80 percentile). The mean South-North gradient over the period is -0.02 $\pm$ 0.02 /10km, and a West-East gradient at -0.002 $\pm$ 0.016 /10km (bottom right panel). The interpolation above both sites is probably unreliable given the context of abrupt orography that surrounds both the Mauna Kea volcano and the Andes pacific coast.  
 \label{fig:A_AE}}
\end{figure}

\end{document}